\documentclass[11pt,oneside,english,reqno]{amsart}
\usepackage{mathpazo}
\usepackage[scaled=0.92]{helvet}
\usepackage{courier}
\usepackage[T1]{fontenc}
\usepackage[latin9]{inputenc}
\usepackage[a4paper]{geometry}
\usepackage{color}
\usepackage{float}
\usepackage{units}
\usepackage{amsbsy}
\usepackage{amstext}
\usepackage{amsthm}
\usepackage{graphicx}
\usepackage{setspace}
\usepackage{wasysym}
\usepackage[authoryear]{natbib}
\PassOptionsToPackage{normalem}{ulem}
\usepackage{ulem}
\makeatletter

\providecommand{\tabularnewline}{\\}
\floatstyle{ruled}
\newfloat{algorithm}{tbp}{loa}
\providecommand{\algorithmname}{Algorithm}
\floatname{algorithm}{\protect\algorithmname}

\numberwithin{equation}{section}
\numberwithin{figure}{section}

\theoremstyle{plain}

\numberwithin{equation}{section}
\usepackage{algorithm,algpseudocode}

\@ifundefined{showcaptionsetup}{}{%
 \PassOptionsToPackage{caption=false}{subfig}}
\usepackage{subfig}
\makeatother

\usepackage{babel}
\begin{document}

\title[Fast Bayesian Whole-Brain fMRI Analysis with Spatial 3D Priors]{Fast Bayesian Whole-Brain fMRI Analysis with Spatial 3D Priors}

\author{Per Sidén, Anders Eklund, David Bolin and Mattias Villani }

\thanks{Per Sidén: \textit{Division of Statistics and Machine Learning, Dept.
of Computer and Information Science, Linköping University, SE-581
83 Linköping, Sweden}. \textit{E-mail: per.siden@liu.se. Corresponding
author.} \\
Anders Eklund: \textit{Division of Statistics and Machine Learning,
Dept. of Computer and Information Science, Division of Medical Informatics,
Dept. of Biomedical Engineering and Center for Medical Image Science
and Visualization (CMIV), Linköping University, SE-581 83 Linköping},
Sweden. \textit{E-mail: anders.eklund@liu.se}. \\
David Bolin: \textit{Division of Mathematical Statistics, Dept. of
Mathematical Sciences, Chalmers and University of Gothenburg, SE-412
96 Göteborg, Sweden. E-mail: david.bolin@chalmers.se.}\\
Mattias Villani: \textit{Division of Statistics and Machine Learning,
Dept. of Computer and Information Science, Linköping University, SE-581
83 Linköping, Sweden}. \textit{E-mail: mattias.villani@liu.se}.}
\begin{abstract}
Spatial whole-brain Bayesian modeling of task-related functional magnetic
resonance imaging (fMRI) is a great computational challenge. Most
of the currently proposed methods therefore do inference in subregions
of the brain separately or do approximate inference without comparison
to the true posterior distribution. A popular such method, which is
now the standard method for Bayesian single subject analysis in the
SPM software, is introduced in \citet{pennyEtAlSpatialPrior2005}.
The method processes the data slice-by-slice and uses an approximate
variational Bayes (VB) estimation algorithm that enforces posterior
independence between activity coefficients in different voxels. We
introduce a fast and practical Markov chain Monte Carlo (MCMC) scheme
for exact inference in the same model, both slice-wise and for the
whole brain using a 3D prior on activity coefficients. The algorithm
exploits sparsity and uses modern techniques for efficient sampling
from high-dimensional Gaussian distributions, leading to speed-ups
without which MCMC would not be a practical option. Using MCMC, we
are for the first time able to evaluate the approximate VB posterior
against the exact MCMC posterior, and show that VB can lead to spurious
activation. In addition, we develop an improved VB method that drops
the assumption of independent voxels a posteriori. This algorithm
is shown to be much faster than both MCMC and the original VB for
large datasets, with negligible error compared to the MCMC posterior.
\end{abstract}

\maketitle

\section{Introduction}

Over the past fifteen years, there has been much work devoted to Bayesian
spatial modeling of task-related functional magnetic resonance imaging
(fMRI) data. The motivation to this line of work has been to develop
an extension to the classical general linear model (GLM) approach
\citep{Friston1995a}. The idea is to replace pre-smoothing of data
and post-correction of multiple hypothesis testing \textendash{} an
approach that was recently shown to be problematic for cluster inference
\citep{Eklund2016} \textendash{} with a proper spatial model. A corner
stone in the field is a series of papers \citep{Penny2003,Penny2005,pennyEtAlSpatialPrior2005,Penny2007,Penny2005a}
upon which the Bayesian spatial single-subject analysis method in
the Statistical Parametric Mapping (SPM) software \citep{SPMsoftware}
is built. The method extends the classical GLM approach to a Bayesian
framework with a spatial Gaussian Markov Random Field (GMRF) prior
on the activity parameters and the temporal noise parameters, encouraging
them to vary smoothly across the brain. The prior is formulated such
that the optimal amount of smoothness can be estimated from the data.
A fast variational Bayes (VB) algorithm is used for inference, but
makes the assumptions that (i) the posterior factorizes over different
types of parameters and (ii) the posterior for the activity parameters
and the temporal noise model parameters factorizes over voxels. Assumption
(i) is validated in the univariate (single-voxel) case in \citet{Penny2003}
by comparing the VB posterior to the exact posterior obtained from
MCMC sampling, but the error from assumption (ii) has to our knowledge
never been properly examined. The VB framework also allows Bayesian
model comparison based on the model evidence lower bound \citep{Penny2007},
but the computation of determinants of spatial precision matrices
limits these types of analyses to be performed slice by slice or to
sub-volumes containing $\apprle10000$ voxels. Even without the model
comparison, the SPM method would be considered too time consuming
for the full 3D brain analysis for most practitioners.

A number of extensions of the SPM method have been developed. \citet{Harrison2008}
replace the stationary spatial prior in \citet{pennyEtAlSpatialPrior2005}
with a non-stationary prior which is more adaptive and find evidence
for this in data from an fMRI study of the auditory system. \citet{Groves2009}
note that the spatial prior in \citet{pennyEtAlSpatialPrior2005}
actually performs simultaneous smoothing and shrinkage of activity
parameters in a rather non-flexible way. They separate these two effects
using a Gaussian Process (GP) prior with a squared exponential kernel
function, and infer the kernel length scale using Evidence Optimization
(EO). The same issue is targeted in \citet{Yue2014}, who use the
GMRF representation of a Matérn field \citep{Lindgren2011} and perform
estimation using Integrated Nested Laplace Approximation (INLA) \citep{Solis-Trapala2009}.

Even though all of these extensions seem like improvements from a
modeling perspective they all struggle with computational complexity
and inference is only performed slice by slice. A computationally
attractive approach towards spatial 3D modeling is to first partition
the brain into sub-volumes or parcels \citep{Thirion2014}, and then
estimate a separate 3D model for each parcel, which can also be done
in parallel, see for example \citet{Harrison2008a,Vincent2010a,Musgrove2015}.
The inability to model dependencies between parcels is a bit unnatural,
however, and \citet{Harrison2008a} notice discontinuities in posterior
estimates along partition boundaries when using their model. Full
3D brain analysis with reasonable speed is achieved by \citet{Harrison2010}
who approximate the zero mean spatial prior in \citet{pennyEtAlSpatialPrior2005}
with a non-zero mean empirical prior, with the cost of no longer having
a true generative model.

An alternative route to a spatial model is to view selecting active
voxels as a variable selection problem, with a spatial prior on the
probability of activation in a certain voxel, see for example \citet{Smith2007}
and \citet{Zhang2014}. A comparison between estimation by VB and
MCMC for this kind of model is available in the recent paper by \citet{Zhang2016a}.
In a related line of work \citep{Vincent2010a,Risser2011,Chaari2013},
voxel activations are modeled using a spatial mixture model of Gaussian
distributions, where one of the mixture components has mean zero,
corresponding to non-active voxels. This framework has the benefit
of allowing for simultaneous estimation of the hemodynamic response
function (HRF), using both MCMC and variational methods which are
applied parcel-wise. Another possibility is to model not only the
activity parameters, but also the noise as spatially dependent (see
for example \citet{Woolrich2004c}), an assumption that seems natural
but which comes with computational trouble because of the spatio-temporal
structure. 

A typical goal with these types of analyses is to compute Posterior
Probability Maps (PPMs) \citep{Friston2003a}, that is, brain maps
showing the marginal probability of activation under a certain stimuli
in each voxel. It is also common that these PPMs are thresholded at
some level, for example at $p>0.9$, to display active voxels or regions.
However, such thresholding implicitly defines a hypothesis test in
each voxel and these multiple tests must be corrected for. Given a
spatially dependent posterior, a natural way to do this is using theory
on excursion sets \citep{Bolin2014a} as in \citet{Yue2014} who define
the term \emph{joint PPM} based on the joint posterior, as opposed
to \emph{marginal PPMs}.

Our paper makes a number of contributions. First, we propose a fast
and practical MCMC algorithm for slice-by-slice and whole brain task-fMRI
analysis with spatial priors on the activity fields and the autoregressive
noise parameters. The algorithm makes efficient use of sparsity and
sampling from high-dimensional Gaussian distributions using preconditioned
conjugate gradient (PCG) methods. These efficiency improvements reduce
the computational complexity by several orders of magnitude. Collectively,
they make it possible to perform posterior whole-brain analysis with
spatial priors on problems where MCMC was previously simply not a
practical option. Second, we develop a very fast VB approach that
maintains the weaker independence assumption (i) in SPM's VB method,
but drops the stronger assumption (ii). Letting go of the assumption
of spatially a posteriori independent voxels is non-trivial from a
computational standpoint, and we employ several numerical techniques
that together make the non-factorized VB approximation a very fast
alternative to MCMC. The approximation errors from this non-factorized
VB are shown to be essentially negligible for practical applications.
Third, we demonstrate that the completely factorized VB in SPM12 can
lead to spurious activations via a hitherto unexplored channel. Factorized
VBs are well known to underestimate posterior variances, but we highlight
and explain why the factorization over voxels can also result in a
quite distorted smoothing of the activations.

The paper is organized as follows. We give a short background on the
spatial model in SPM and the VB method used there to estimate the
model. We then introduce the MCMC algorithm and the improved VB method
and put extra emphasis on how to make these methods computationally
efficient. In the next section we show results for simulated and real
experiment data, and illustrate how the parameter estimates and PPMs
differ between the different methods. The speed and convergence properties
across methods are also compared. The last section contains a discussion
and our conclusions. Derivation of results and implementation details
can be found in the appendices.

The new methods presented in this article have been implemented as
an extension to the SPM software, available at \textcolor{blue}{\uline{http://www.fil.ion.ucl.ac.uk/spm/ext/\#BFAST3D}}.

\section{Background\label{sec:Background}}

We will consider single-subject fMRI-data containing $T$ volumes
with $N$ voxels ordered in a $T\times N$ matrix $\mathbf{Y}$. The
experiment is described by the $T\times K$ design matrix $\mathbf{X}$,
with $K$ regressors. The model in \citet{Penny2007} is written as
\begin{equation}
\mathbf{Y}=\mathbf{X}\mathbf{W}+\mathbf{E},\label{eq:Pennymodel}
\end{equation}
where $\mathbf{W}$ is a $K\times N$ matrix of regression coefficients
and \textbf{$\mathbf{E}$} is a $T\times N$ matrix of error terms.
Since this is the same model that will be considered here, we will
use the same notation throughout, unless stated otherwise. Voxel-specific,
normally distributed $P$th order AR models are assumed for the error
terms, but for a clearer presentation here we will consider the special
case $P=0$ and handle the more general case in Appendix \ref{Appendix full conditionals derivations}.
The likelihood then becomes{\footnotesize{} }
\begin{eqnarray}
p\left(\mathbf{Y}|\mathbf{W},\boldsymbol{\lambda}\right) & = & \prod_{n=1}^{N}\mathcal{N}\left(\mathbf{Y}_{\cdot,n};\mathbf{X}\mathbf{W}_{\cdot,n},\lambda_{n}^{-1}\mathbf{I}_{T}\right),\label{eq:likelihood}
\end{eqnarray}
with $\mathbf{Y}_{\cdot,n}$ and $\mathbf{W}_{\cdot,n}$ denoting
the $n$th column of $\mathbf{Y}$ and $\mathbf{W},$ $\lambda_{n}$
as the noise precision of voxel $n$ and $\mathbf{I}_{T}$ a $T\times T$
identity matrix. The likelihood factorizes over voxels, which is an
assumption of non-spatial measurement noise that is made because the
opposite assumption would be very computationally challenging. Instead,
the spatial part of the model will enter via the prior on the regression
coefficients
\begin{eqnarray}
\mathbf{W}_{k,\cdot}^{\prime}|\alpha_{k} & \sim & \mathcal{N}\left(0,\alpha_{k}^{-1}\mathbf{D}_{w}^{-1}\right),\label{eq:Wk prior}\\
p\left(\mathbf{W}|\boldsymbol{\alpha}\right) & = & \prod_{k=1}^{K}p\left(\mathbf{W}_{k,\cdot}^{\prime}|\alpha_{k}\right),\nonumber 
\end{eqnarray}
where $\mathbf{W}_{k,\cdot}^{\prime}$ denotes the transposed $k$th
row of $\mathbf{W}$, $\mathbf{D}_{w}$ is a fixed spatial $N\times N$
precision matrix and $\boldsymbol{\alpha}=\left(\alpha_{1},\ldots,\alpha_{K}\right)^{\prime}$
are hyperparameters to be estimated from the data. There are several
possible choices for $\mathbf{D}_{w}$, but we will here focus on
the SPM12 default choice, the unweighted graph-Laplacian (UGL) (called
$\mathbf{L}$ in \citet{pennyEtAlSpatialPrior2005}) which for each
voxel has the number of adjacent voxels on the diagonal and $\mathbf{D}_{w}(i,j)=-1$
if $i$ and $j$ are adjacent. For voxels in the interior part of
the brain we will thus have $6$'s on the diagonal when modeling the
whole 3D brain, and $4$'s when modeling each 2D slice separately.
The main focus in \citet{pennyEtAlSpatialPrior2005} is on a different
prior defined as $\mathbf{D}_{w}=\mathbf{L}^{\prime}\mathbf{L}$.
It is straightforward to use that prior also in our framework, but
our experience is that it leads to too smooth posteriors, which is
probably the reason why it is not the default option in SPM12. It
also leads to slower inference since $\mathbf{D}_{w}$ is less sparse
when using this prior. The assumption of a sparse $\mathbf{D}_{w}$
is the key to fast inference for this type of model using any method.

The hyperparameter $\alpha_{k}$ will be estimated for each regressor
to put the appropriate weight on the prior, depending on what smoothness
is supported by the data. A higher $\alpha_{k}$ brings the regression
coefficient in each voxel closer to the mean of the coefficients in
neighboring voxels (more smoothness), and globally all coefficients
closer to zero (more shrinkage). The precision parameters $\alpha_{k}$
and $\lambda_{n}$ are assigned conjugate Gamma priors. For all the
details on the generative model and the priors, see Figure 1 and Appendix
A in \citet{Penny2007} which also gives the values of the prior parameters
that are default in SPM12 which are also used here (except $r_{1}=10000$
in SPM12). That is, we use $q_{1}=u_{1}=10$, $r_{1}=10000$ and $q_{2}=r_{2}=u_{2}=0.1$.

The VB algorithm for inference presented in \citet{pennyEtAlSpatialPrior2005}
makes two independence assumptions for the joint posterior distribution.
Firstly, the posterior is assumed to factorize over the different
kinds of parameters, that is
\begin{equation}
q\left(\mathbf{W},\boldsymbol{\alpha},\boldsymbol{\lambda}|\mathbf{Y}\right)=q\left(\mathbf{W}|\mathbf{Y}\right)q\left(\boldsymbol{\alpha}|\mathbf{Y}\right)q\left(\boldsymbol{\lambda}|\mathbf{Y}\right),\label{eq:factorized posterior}
\end{equation}
with $q$ denoting VB posteriors. Secondly, the regression parameters
(and the AR-parameters in the general case) are assumed to factorize
over voxels
\begin{equation}
q\left(\mathbf{W}|\mathbf{Y}\right)=\prod_{n=1}^{N}q\left(\mathbf{W}_{\cdot,n}|\mathbf{Y}\right).\label{eq:factorized posterior 2}
\end{equation}
The second assumption is possibly the strongest and most counter-intuitive
since it is clear that the spatial prior will generate dependence
between voxels in the posterior. In the following section we will
present an efficient MCMC (Gibbs) algorithm that performs inference
without any of these two assumptions and an improved VB algorithm
that drops only the second one. For an introduction to MCMC and VB,
see \citet{Penny2003}, who introduce these methods for the one-voxel
case.

\section{Theory}

\citet{pennyEtAlSpatialPrior2005} motivate the second independence
assumption by noting that the posterior distribution for $\mathbf{W}$
will otherwise have a full covariance matrix of size $KN\times KN$,
which is prohibitively large. This is certainly true, and our algorithms
are therefore designed to never compute the covariance matrix explicitly,
but work with precision matrices instead. The posterior precision
matrix is also of size $KN\times KN$, but it is sparse and can therefore
be stored and used for computations quite cheaply. For example, the
full conditional posterior distribution for $\mathbf{w}_{r}=vec\left(\mathbf{W}^{\prime}\right)$
will be multivariate normal and can be characterized by (see Appendix
\ref{Appendix full conditionals derivations} for the derivation)
\begin{eqnarray}
p\left(\mathbf{w}_{r}|\mathbf{Y},\boldsymbol{\alpha},\boldsymbol{\lambda}\right) & \propto & \exp\left(-\frac{1}{2}\mathbf{w}_{r}^{\prime}\tilde{\mathbf{B}}\mathbf{w}_{r}+\mathbf{b}_{w}^{\prime}\mathbf{w}_{r}\right),\label{eq:wr full conditional}\\
\mathbf{b}_{w} & = & vec\left(diag\left(\boldsymbol{\lambda}\right)\mathbf{Y}^{\prime}\mathbf{X}\right),\nonumber \\
\tilde{\mathbf{B}} & = & \mathbf{X}^{\prime}\mathbf{X}\otimes diag\left(\boldsymbol{\lambda}\right)+diag\left(\boldsymbol{\alpha}\right)\otimes\mathbf{D}_{w},\nonumber 
\end{eqnarray}
where $\tilde{\mathbf{B}}$ is a precision matrix which will maximally
have $K+6$ non-zero elements in any row when the UGL prior is used
in 3D. In other words, $\mathbf{w}_{r}|\mathbf{Y},\boldsymbol{\lambda},\boldsymbol{\alpha}\sim\mathcal{N}\left(\tilde{\mathbf{B}}^{-1}\mathbf{b}_{w},\tilde{\mathbf{B}}^{-1}\right).$

As all the matrix inversions are avoided, the computational bottleneck
of the algorithms developed in this paper will instead be to generate
random samples from this and similar normal distributions, and to
solve equation systems of the form $\tilde{\mathbf{B}}\mathbf{x}=\mathbf{b}$,
given some $KN\times1$ vector $\mathbf{b}$. We will approach these
problems in two ways, using Cholesky decomposition-based exact methods
and preconditioned conjugate gradient (PCG) approximate methods.

\subsection{GMRF sampling}

~\\
\citet{Isham2004} give a nice introduction to inference in latent
Gaussian spatial models with sparse precision matrices, so called
Gaussian Markov Random Fields (GMRFs), and also provide a good list
of historical references, for example \citet{Besag1974} and \citet{Woods1972}.
In particular, they give computationally effective algorithms for
sampling and equation solving for GMRFs based on first computing the
Cholesky factor of the precision matrix, that is for a precision matrix
$\tilde{\mathbf{B}}$ compute a lower triangular matrix $\mathbf{L}$
such that $\mathbf{L}\mathbf{L}^{\prime}=\tilde{\mathbf{B}}.$ Using
so called reordering methods, such as the approximate minimum degree
permutation \citep{Amestoy1996}, one can find a way to reorder the
rows and columns of a sparse $\tilde{\mathbf{B}}$ such that $\mathbf{L}$
will be reasonably sparse as well, which will be important for speed.
We can sample from the distribution in equation (\ref{eq:wr full conditional})
and calculate $E\left(\mathbf{w}_{r}|\mathbf{Y},\boldsymbol{\alpha},\boldsymbol{\lambda}\right)=\boldsymbol{\mu}_{w}=\tilde{\mathbf{B}}^{-1}\mathbf{b}_{w}$
using Algorithm \ref{Alg:Cholesky sampling}. A key notion here is
that, algorithmically, solving equations involving $\mathbf{L}$ or
$\mathbf{L}^{\prime}$ (called forward or backward substitution) is
much faster than solving equations involving $\tilde{\mathbf{B}}$
directly. In practice, this algorithm works well for GMRFs of dimension
up to $\approx10000$, but then starts to become too slow for practical
use, the bottleneck being the Cholesky decomposition. This means that
it can be used for slice- or parcel-wise inference in many cases,
but not for whole 3D brain inference.

\citet{Papandreou2010} offer a method to sample from the posterior
for $\mathbf{W}$ that avoids the Cholesky decomposition, see Algorithm
\ref{Alg:PCG sampling}. To use it in our setting we first have to
rewrite the prior density for $\mathbf{W}_{k,\cdot}^{\prime}$ according
to
\begin{equation}
p\left(\mathbf{W}_{k,\cdot}^{\prime}|\alpha_{k}\right)\propto\mathcal{N}\left(\sqrt{\alpha_{k}}\mathbf{G}_{w}\mathbf{W}_{k,\cdot}^{\prime};0,\mathbf{I}_{N_{\mathbf{G}_{w}}}\right)\propto\exp\left(-\frac{1}{2}\mathbf{W}_{k,\cdot}\alpha_{k}\mathbf{G}_{w}^{\prime}\mathbf{G}_{w}\mathbf{W}_{k,\cdot}^{\prime}\right),\label{eq:PCG prior}
\end{equation}
for some $N_{\mathbf{G}_{w}}\times N$ matrix $\mathbf{G}_{w}$. For
this prior to be equal to the one in equation (\ref{eq:Wk prior}),
$\mathbf{G}_{w}$ needs to be chosen such that $\mathbf{D}_{w}=\mathbf{G}_{w}^{\prime}\mathbf{G}_{w}$.
Fortunately, many priors are naturally specified through $\mathbf{G}_{w}$
directly. In our case, we construct $\mathbf{G}_{w}$ based on the
interpretation of the UGL as a priori saying that the differences
between adjacent voxels are i.i.d. normal, that is
\begin{equation}
W_{k,i}-W_{k,j}\overset{^{iid}}{\sim}\mathcal{N}\left(0,\alpha_{k}^{-1}\right),\label{eq:indep increments}
\end{equation}
for all adjacent $i$ and $j$. Thus, we can construct $\mathbf{G}_{w}$
as having one row for every pair of adjacent voxels $i$ and $j$
with $1$ in column $i$ and $-1$ in column $j$. A similar construction
is possible for the main prior in \citet{pennyEtAlSpatialPrior2005}.
In cases where the prior is instead specified through $\mathbf{D}_{w}$,
one way to obtain a $\mathbf{G}_{w}$ is always available as the Cholesky
factor of $\mathbf{D}_{w}$, as long as this is computable. Further,
we construct $\mathbf{B}_{data}=diag\left(\boldsymbol{\lambda}\right)\otimes\mathbf{X}^{\prime}\mathbf{X}$
and $\mathbf{L}_{data}$ as its Cholesky factor which is cheap to
compute since $\mathbf{B}_{data}$ will be banded with bandwidth $K$
and also block diagonal. We define $\mathbf{H}_{w}$ as in \citet{Penny2007}
as the permutation matrix such that $vec\left(\mathbf{W}\right)=\mathbf{H}_{w}vec\left(\mathbf{W}^{\prime}\right)$. 

The last piece of the second sampling method is to use PCG for solving
equations of the form $\tilde{\mathbf{B}}\mathbf{x}=\mathbf{b}$ approximately
\citep{barrett1994templates,Manteuffel1980}, with a computationally
cheap incomplete Cholesky pre-conditioner $\mathbf{M}$. The efficiency
of the PCG method increases if given a starting value $\mathbf{x}^{start}$
that is close to the solution. The PCG algorithm will iterate until
the relative residual $\left\Vert \tilde{\mathbf{B}}\mathbf{x}-\mathbf{b}\right\Vert /\left\Vert \mathbf{b}\right\Vert $
becomes less than a user specified tolerance level $\delta$, e.g.
$10^{-8}$, why PCG can be set to approximate the true solution arbitrarily
well. This means that the approximation error that comes from PCG
will be negligible in practice, which is exemplified in Figure \ref{fig:RMSEBySize}
below, where the difference in posterior mean for the MCMC method
using $\delta=10^{-6}$ and $\delta=10^{-8}$ is very close to zero.
PCG can be used both as in Algorithm \ref{Alg:PCG sampling} for sampling
or simply to compute the mean $\boldsymbol{\mu}_{w}=\tilde{\mathbf{B}}^{-1}\mathbf{b}_{w}$.

\begin{algorithm}[H]
\begin{algorithmic}[1]
\Require $\tilde{\mathbf{B}},\mathbf{b}_{w}$
\vspace{1mm}
\State Compute reordering based on $\tilde{\mathbf{B}}$ and reorder $\tilde{\mathbf{B}}$ and $\mathbf{b}_{w}$ accordingly
\vspace{1mm}
\State Compute $\mathbf{L}$ as the Cholesky factor of $\tilde{\mathbf{B}}$
\vspace{1mm}
\State Solve $\mathbf{L} \mathbf{x}=\mathbf{b}_w $
\vspace{1mm}
\State Solve $\mathbf{L}^{\prime}\boldsymbol{\mu}_{w}=\mathbf{x}$
\vspace{1mm}
\State Sample $\mathbf{z}\sim N(0,\mathbf{I}_{KN})$
\vspace{1mm}
\State Solve $\mathbf{L}^{\prime}\mathbf{v}=\mathbf{z}$
\vspace{1mm}
\State Compute $\mathbf{w}_r=\boldsymbol{\mu}_w+\mathbf{v}$
\vspace{1mm}
\State Reorder $\mathbf{w}_r$ and $\boldsymbol{\mu}_w$ using the inverse reordering computed in step 1
\vspace{1mm}
\State \Return $\mathbf{w}_r,\boldsymbol{\mu}_w$
\end{algorithmic}\caption{Cholesky based sampling from $p\left(\mathbf{w}_{r}|\mathbf{Y},\boldsymbol{\alpha},\boldsymbol{\lambda}\right)$\label{Alg:Cholesky sampling}}
\end{algorithm}

\begin{algorithm}[H]
\begin{algorithmic}[1]
\Require $\tilde{\mathbf{B}},\mathbf{b}_{w},\mathbf{G}_w,\mathbf{B}_{data},\mathbf{w}_r^{start}$
\vspace{1mm}
\State Compute $\mathbf{L}_{data}$ as the Cholesky factor of $\mathbf{B}_{data}$
\vspace{1mm}
\State Sample $\mathbf{z}_1\sim \mathcal{N}(0,\mathbf{I}_{N_{G_w}})$
\vspace{1mm}
\State Sample $\mathbf{z}_2\sim \mathcal{N}(0,\mathbf{I}_{KN})$
\vspace{1mm}
\State Compute $\mathbf{b}=\left(\underset{k\in\left\{ 1,\ldots,K\right\} }{blkdiag}\left[\sqrt{\alpha_{k}}\mathbf{G}_w\right]\right)^{\prime}\mathbf{z}_1+\mathbf{H}_w^{\prime}\mathbf{L}_{data}\mathbf{H}_w \mathbf{z}_2+\mathbf{b}_{w}$
\vspace{1mm}
\State Compute reordering based on $\tilde{\mathbf{B}}$ and reorder $\tilde{\mathbf{B}}$ and $\mathbf{b}$ accordingly
\vspace{1mm}
\State Compute $\mathbf{M}$ as the Incomplete Cholesky factor of $\tilde{\mathbf{B}}$
\vspace{1mm}
\State Solve $\tilde{\mathbf{B}}\mathbf{w}_r=\mathbf{b}$ approximately using PCG with preconditioner $\mathbf{M}$ and starting value $\mathbf{w}_r^{start}$
\vspace{1mm}
\State Reorder $\mathbf{w}_r$ using the inverse reordering computed in step 5
\vspace{1mm}
\State \Return $\mathbf{w}_r$
\end{algorithmic}\caption{PCG based sampling from $p\left(\mathbf{w}_{r}|\mathbf{Y},\boldsymbol{\alpha},\boldsymbol{\lambda}\right)$\label{Alg:PCG sampling}}
\end{algorithm}

\begin{figure}[H]
\includegraphics[width=0.5\linewidth]{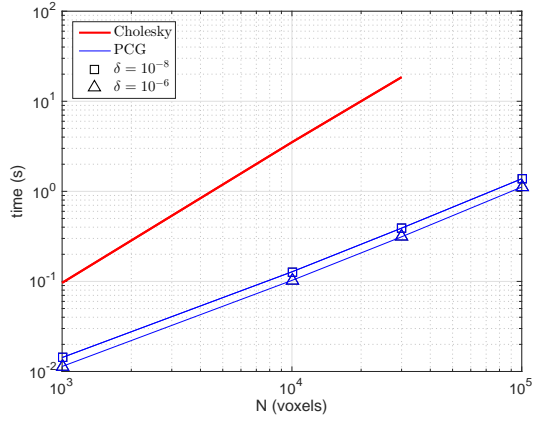}

\caption{Average sampling times using Cholesky and PCG based sampling for GMRFs
of size $KN$ for the simulated data $\left(K=5\right)$, with two
different PCG tolerance levels $\delta$. In each case the processing
time was computed as an average over 100 samples. Note that the presented
PCG timings can be reduced by an additional factor in the range $[2,15]$
when a good starting value is available. The Cholesky algorithm runs
out memory for $N=10^{5}$. \label{fig:samplingTiming} }
\end{figure}

Figure \ref{fig:samplingTiming} shows the average time it takes to
produce a single sample from the conditional posterior of $\mathbf{W}$
for the simulated data presented in Section \ref{subsec:Simulated-data}
$\left(K=5\right)$ for the Cholesky and PCG based sampling algorithms,
as a function of the number of voxels. The Cholesky algorithm runs
out of memory (32GB RAM) for $N=10^{5}$ (which is another issue for
the Cholesky approach), but extending the seemingly linear behavior
on the log-log-scale indicates that the PCG method would be roughly
100 times faster in this case. By providing a good starting value,
as in the SVB algorithm presented below, we have observed the PCG
timings to decrease with an additional factor in the range $\left[2,15\right]$,
a factor that increases as the algorithm converges.

\subsection{MCMC algorithm}

~\\
In order to evaluate the true posterior of the model without any independence
assumptions between parameters, we develop an algorithm for MCMC sampling.
Since all priors are conjugate, we obtain closed form expressions
for all full conditionals and can therefore perform Gibbs sampling.
The full conditional posterior for $\mathbf{W}$ is given in (\ref{eq:wr full conditional})
and the full conditionals for $\boldsymbol{\lambda}$ and $\boldsymbol{\alpha}$
are given by
\begin{eqnarray}
\lambda_{n}|\mathbf{Y},\mathbf{W},\boldsymbol{\alpha} & \sim & Ga\left(\tilde{u}_{1n},\tilde{u}_{2}\right),\label{eq:full conditionals}\\
\alpha_{k}|\mathbf{Y},\mathbf{W},\boldsymbol{\lambda} & \sim & Ga\left(\tilde{q}_{1k},\tilde{q}_{2}\right),\nonumber 
\end{eqnarray}
with
\begin{eqnarray}
\frac{1}{\tilde{u}_{1n}} & = & \frac{1}{2}\left(\mathbf{Y}_{\cdot,n}^{\prime}\mathbf{Y}_{\cdot,n}-2\mathbf{Y}_{\cdot,n}^{\prime}\mathbf{X}\mathbf{W}_{\cdot,n}+\mathbf{W}_{\cdot,n}^{\prime}\mathbf{X}^{\prime}\mathbf{X}\mathbf{W}_{\cdot,n}\right)+\frac{1}{u_{1}},\label{eq:posterior parameters}\\
\tilde{u}_{2} & = & \frac{T}{2}+u_{2},\nonumber \\
\frac{1}{\tilde{q}_{1k}} & = & \frac{1}{2}\mathbf{W}_{k,\cdot}\mathbf{D}_{w}\mathbf{W}_{k,\cdot}^{\prime}+\frac{1}{q_{1}},\nonumber \\
\tilde{q}_{2} & = & \frac{N}{2}+q_{2}.\nonumber 
\end{eqnarray}
See Appendix \ref{Appendix full conditionals derivations} for the
derivation of these and the corresponding full conditionals in the
$P>0$ case. 

The Gibbs algorithm returns $N_{iter}$ samples from the joint posterior
of all parameters, which can be used for posterior inference about
any subset of parameters. For example, given samples $\mathbf{W}^{\left(1:N_{iter}\right)}$
we can compute the marginal PPM for any $K\times1$ contrast vector
$\mathbf{c}$ as
\begin{equation}
P\left(\mathbf{c}^{\prime}\mathbf{W}_{\cdot,n}>\gamma|\mathbf{Y}\right)\approx\frac{1}{N_{iter}}\sum_{j=1}^{N_{iter}}I\left(\mathbf{c}^{\prime}\mathbf{W}_{\cdot,n}^{\left(j\right)}>\gamma\right),\label{eq:MCMCPPM}
\end{equation}
for voxel $n$ and some activity threshold $\gamma$. Furthermore,
since the MCMC posterior does not factorize over voxels, it is also
meaningful and straightforward to compute the joint probability of
activation for any set of voxels $E$ as
\begin{equation}
P\left(\mathbf{c}^{\prime}\mathbf{W}_{\cdot,E}>\gamma\boldsymbol{1}_{\left|E\right|}^{\prime}|\mathbf{Y}\right)\approx\frac{1}{N_{iter}}\sum_{j=1}^{N_{iter}}I\left(\mathbf{c}^{\prime}\mathbf{W}_{\cdot,E}^{\left(j\right)}>\gamma\boldsymbol{1}_{\left|E\right|}^{\prime}\right),\label{eq:MCMCjointPPM}
\end{equation}
where $\boldsymbol{1}_{\left|E\right|}$ is a vector of ones of length
$\left|E\right|$. Using the theory on excursion sets developed by
\citet{Bolin2014a}, we can thereby compute the joint PPMs introduced
in \citet{Yue2014} that avoid the problem of multiple hypothesis
testing.

\subsection{Spatial VB\label{subsec:Spatial-VB}}

~\\
Out of the two posterior independence assumptions in \citet{pennyEtAlSpatialPrior2005},
we view the second one as the strongest, that is the assumption that
the posterior for $\mathbf{W}$ factorizes over voxels. The developed
MCMC algorithm relieves us from both independence assumptions, but
has its limitations in terms of speed and memory. We therefore seek
to develop an improved VB algorithm that maintains the efficiency
gain from the first assumption of independence between the different
types of parameters, but drops the second assumption and models the
joint posterior of $\mathbf{W}$. We will refer to this algorithm
as Spatial Variational Bayes (SVB) and to SPM's factorized VB algorithm
as Independent Variational Bayes (IVB).

The SVB posterior will be computed iteratively, just as in the IVB
algorithm, for one parameter at a time given the approximate posterior
of the others. If we denote $\boldsymbol{\theta}=\left\{ \mathbf{W},\boldsymbol{\lambda},\boldsymbol{\alpha}\right\} $,
then \citep{Bishop2006}
\begin{equation}
\log q\left(\theta_{j}\right)=E_{\theta_{-j}}\left[\log p\left(\mathbf{Y},\boldsymbol{\theta}\right)\right]+const,\label{eq:Bishops vb posterior}
\end{equation}
with the expectation taken with respect to the VB posterior $q\left(\theta_{-i}\right)$.
This means that for $q\left(\mathbf{W}\right)$ we get (see Appendix
\ref{Appendix NFVB derivations})
\begin{eqnarray}
\log q\left(\mathbf{w}_{r}\right) & = & E_{\lambda,\boldsymbol{\alpha}}\left[\log p\left(\mathbf{Y}|\mathbf{W},\boldsymbol{\lambda}\right)+\log p\left(\mathbf{W}|\boldsymbol{\alpha}\right)\right]+const\label{eq:NFVB w posterior}\\
 & = & -\frac{1}{2}\mathbf{w}_{r}^{\prime}\tilde{\mathbf{B}}^{SVB}\mathbf{w}_{r}+\mathbf{b}_{w}^{SVB\prime}\mathbf{w}_{r}+const,\nonumber \\
\mathbf{b}_{w}^{SVB} & = & vec\left(diag\left(E_{\boldsymbol{\lambda}}\left[\boldsymbol{\lambda}\right]\right)\mathbf{Y}^{\prime}\mathbf{X}\right),\nonumber \\
\tilde{\mathbf{B}}^{SVB} & = & \mathbf{X}^{\prime}\mathbf{X}\otimes diag\left(E_{\boldsymbol{\lambda}}\left[\boldsymbol{\lambda}\right]\right)+diag\left(E_{\boldsymbol{\alpha}}\left[\boldsymbol{\alpha}\right]\right)\otimes\mathbf{D}_{w},\nonumber 
\end{eqnarray}
that is, we get exactly the same expression as for the full conditionals
only the values for $\boldsymbol{\lambda}$ and $\boldsymbol{\alpha}$
are replaced by their expectations with respect to their variational
posteriors $q\left(\boldsymbol{\lambda}\right)$ and $q\left(\boldsymbol{\alpha}\right)$.
This expression is simple because $\boldsymbol{\lambda}$ and $\boldsymbol{\alpha}$
enter linearly, which will not be the case in general. For $q\left(\boldsymbol{\alpha}\right)$
we get (note that the likelihood $p\left(\mathbf{Y}|\mathbf{W},\boldsymbol{\lambda}\right)$
does not depend on $\boldsymbol{\alpha}$) 
\begin{eqnarray}
\log q\left(\boldsymbol{\alpha}\right) & = & E_{\mathbf{W},\boldsymbol{\lambda}}\left[\log p\left(\mathbf{Y}|\mathbf{W},\boldsymbol{\lambda}\right)+\log p\left(\mathbf{W}|\boldsymbol{\alpha}\right)+\log p\left(\boldsymbol{\alpha}\right)\right]+const\label{eq:NFVB-alpha}\\
 & = & \left(\frac{N}{2}+q_{2}-1\right)\sum_{k=1}^{K}\log\alpha_{k}-\sum_{k=1}^{K}\alpha_{k}\left[\frac{1}{2}E_{\mathbf{W}}\left[\mathbf{W}_{k,\cdot}\mathbf{D}_{w}\mathbf{W}_{k,\cdot}^{\prime}\right]+\frac{1}{q_{1}}\right]+const.\nonumber 
\end{eqnarray}
So, just as for the full conditionals, $\alpha_{k}$ will be Gamma
distributed a posteriori with parameters
\begin{eqnarray}
\frac{1}{\tilde{q}_{1k}^{SVB}} & = & \frac{1}{2}E_{\mathbf{W}}\left[\mathbf{W}_{k,\cdot}\mathbf{D}_{w}\mathbf{W}_{k,\cdot}^{\prime}\right]+\frac{1}{q_{1}},\label{eq:NFVB-alpha-parameters}\\
\tilde{q}_{2}^{SVB} & = & \frac{N}{2}+q_{2}.\nonumber 
\end{eqnarray}
The problem here is the expectation of the quadratic form
\begin{equation}
E_{\mathbf{W}}\left[\mathbf{W}_{k,\cdot}\mathbf{D}_{w}\mathbf{W}_{k,\cdot}^{\prime}\right]=E_{\mathbf{W}}\left[\mathbf{W}_{k,\cdot}\right]\mathbf{D}_{w}E_{\mathbf{W}}\left[\mathbf{W}_{k,\cdot}^{\prime}\right]+tr\left(\mathbf{D}_{w}Cov\left[\mathbf{W}_{k,\cdot},\mathbf{W}_{k,\cdot}\right]\right),\label{eq:quad-form-expectation}
\end{equation}
where the second term requires inversion (or at least partial inversion)
of the posterior precision for $\mathbf{W}$, $\tilde{\mathbf{B}}^{SVB}$,
which is computationally infeasible in the general 3D case. To avoid
this, we adopt a Monte Carlo (MC) sampling based approach to compute
the expectation. The PCG sampling method provides an efficient way
to generate a number of ($N_{s}$) samples $\mathbf{W}^{\left(1:N_{s}\right)}$
from the VB posterior $q(\mathbf{W})$, by simply replacing $\tilde{\mathbf{B}}$
and $\mathbf{b}_{w}$ in Algorithm \ref{Alg:PCG sampling} with $\tilde{\mathbf{B}}^{SVB}$
and $\mathbf{b}_{w}^{SVB}$. These samples can be used to approximate
the expectation as
\begin{equation}
E_{\mathbf{W}}\left[\mathbf{W}_{k,\cdot}\mathbf{D}_{w}\mathbf{W}_{k,\cdot}^{\prime}\right]\approx\frac{1}{N_{s}}\sum_{j=1}^{N_{s}}\mathbf{W}_{k,\cdot}^{\left(j\right)}\mathbf{D}_{w}\mathbf{W}_{k,\cdot}^{(j)\prime}.\label{eq:MC approx}
\end{equation}
Similar MC approximations will be used in the VB update equations
of $\boldsymbol{\lambda}$ and for the other parameters when $P>0$,
see Appendix \ref{Appendix NFVB derivations}.

The SVB algorithm is similar to the MCMC algorithm in that the computational
bottleneck will be the sampling of $\mathbf{W}$, but there are some
important differences. The MCMC algorithm runs for a large number
of iterations ($N_{iter}$, thousands) producing one sample from $p\left(\mathbf{W}|\mathbf{Y},\boldsymbol{\lambda},\boldsymbol{\alpha}\right)$
in each iteration. For SVB it is enough to run for a much smaller
number of iterations (tens), but each iteration draws a larger number
of samples ($N_{s}$, tens or hundreds) from $q\left(\mathbf{W}\right).$
When PCG based sampling is used SVB is advantageous because (i) the
same pre-conditioner $\mathbf{M}$ can be used for all samples in
each VB iteration (ii) the same random seeds (the same $\mathbf{z}_{1}$
and $\mathbf{z}_{2}$ in Algorithm \ref{Alg:PCG sampling}) can be
used across VB iterations, making the previous iteration samples very
good starting values for the PCG (iii) since the samples are independent,
the sampling can be fully parallelized within each iteration by running
Algorithm \ref{Alg:PCG sampling} on separate cores. All these points
contribute to that SVB can be run much faster than MCMC in general,
with the price being the first assumption of independence between
different kinds of parameters.

The MC approximation introduced in equation (\ref{eq:MC approx})
adds a stochastic approximation error to the already approximate VB
posterior. Figure \ref{fig:RMSEBySize} below quantifies the size
of this error with respect to the number of samples $N_{s}$ on simulated
data. Convergence results for stochastic VB methods \citep{Kingma2013,Gunawan2016}
and for stochastic variants of the related expectation-maximization
(EM) algorithm \citep{Chan1995,Delyon1999} are available in the literature.
However, these results do not apply to our setting since they build
on repeated sampling, while the SVB algorithm presented here only
draws $\mathbf{z}_{1}$ and $\mathbf{z}_{2}$ in Algorithm \ref{Alg:PCG sampling}
in the first iteration and then re-uses those same random numbers
at the subsequent iterations. Re-using the random numbers speeds up
SVB (typically by a factor between $2$ and $15$) since it allows
us to use good PCG starting values, and the additional approximation
error resulting from fixed random numbers is small in our applications.
See Appendix \ref{appendix:convergence} for more details about the
convergence.

In theory, all SVB posterior statistics, such as PPMs, can be computed
from the approximate posterior $q\left(\mathbf{W}\right)$ in equation
(\ref{eq:NFVB w posterior}). However, since it is parametrized using
the precision matrix even basic marginal statistics, as variances
and posterior probabilities, are not trivially obtained since this
requires the inversion (or at least the Cholesky factor) of the precision
matrix. We can instead use the sample variance of the samples $\mathbf{W}^{\left(1:N_{s}\right)}$
to get a fast approximation \citep{Papandreou2011}, which can be
used to compute the marginal PPMs in each voxel. For contrast PPMs,
we similarly use the sample covariance matrix for each voxel. While
a small value of $N_{s}$ $\left(\leq100\right)$ seems sufficient
for convergence of the SVB algorithm, covariance estimates based on
the same number of samples will be quite noisy (see Figure \ref{fig:MeanStdMCMCvsSPMsVB}
below). A straightforward strategy to reduce the noise would be to
generate additional samples from $q\left(\mathbf{W}\right)$ in a
post-processing step, to further improve the covariance estimates.
Such a step could be time consuming, however, and we are currently
working on a different, more efficient way to compute covariances
for a given sparse precision matrix.

\section{Results\label{sec:Results}}

In this section we present results comparing the three methods (IVB,
SVB and MCMC) on the same data. As MCMC is exact (in the sense of
being simulation consistent with a small and controllable error),
we can view this as the ground truth when evaluating the other methods.
IVB is run using SPM12. The SVB algorithm is implemented as an extension
to the original IVB algorithm by manipulating the original SPM12 Matlab
code, while the MCMC algorithm is implemented in a separate Matlab
function. The overhead time is low compared to the main computation
steps for all three methods, so all timing comparisons should be considered
fair. We perform the comparison on simulated data with the focus on
computational efficiency as a function of data size, and on two different
real data sets with the focus on the resulting posteriors and PPMs. 

For any comparison to be fair, both regarding computing time and estimated
posteriors, we need the different algorithms to reach the same level
of convergence. In Appendix \ref{appendix:convergence} we discuss
some details on the convergence of the different methods and show
that the SPM12 default setting of $4$ VB iterations is usually not
sufficient. In the results below, all methods including IVB are run
until convergence. In Appendix \ref{appendix:convergence} we also
provide some practical details about the implementation of respective
method and about the computers used to perform the analyses.

\subsection{Simulated data\label{subsec:Simulated-data}}

\begin{figure}[H]
\subfloat[\label{fig:timingBySize}]{\includegraphics[width=0.5\linewidth]{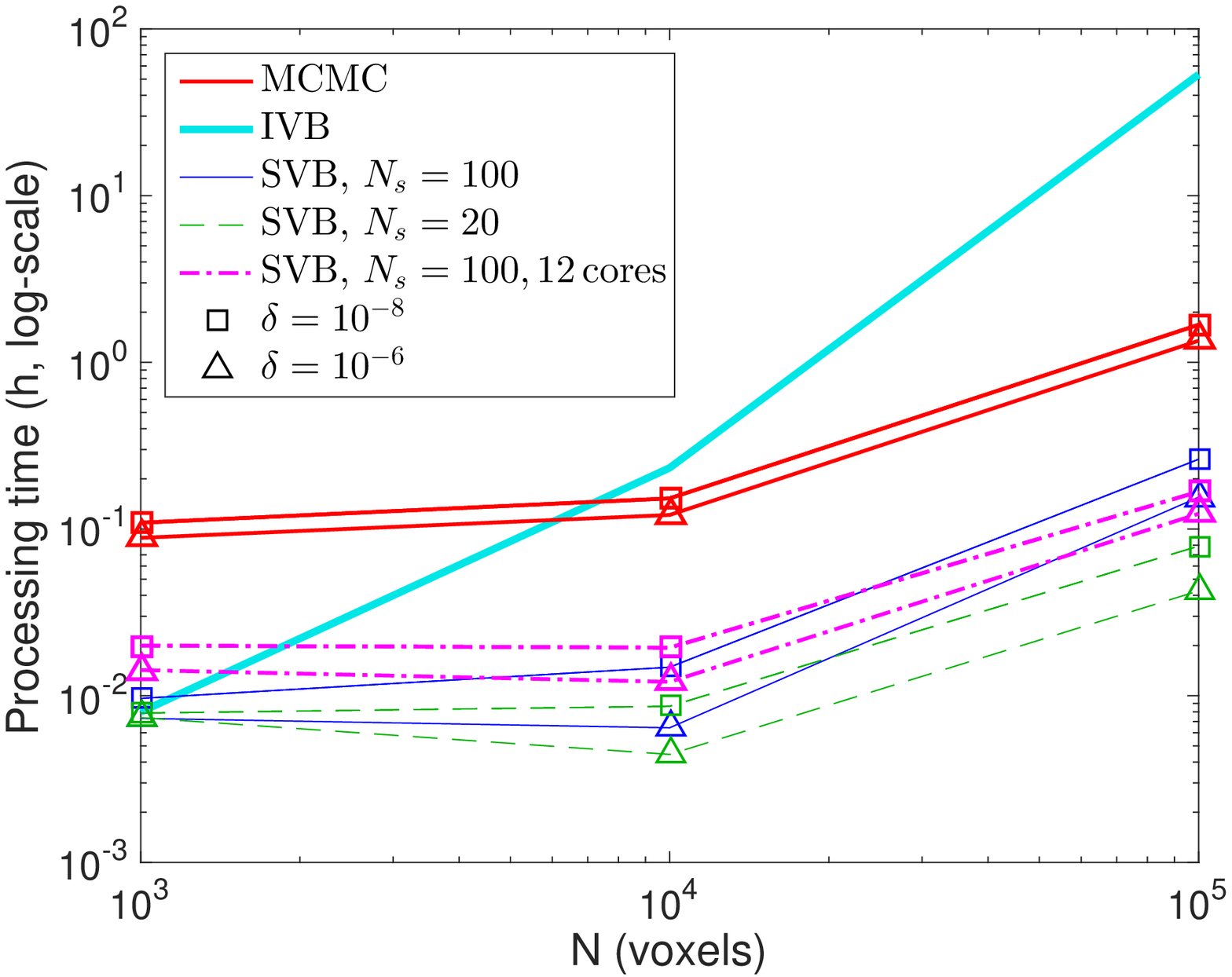}}\subfloat[\label{fig:RMSEBySize}]{\includegraphics[width=0.5\linewidth]{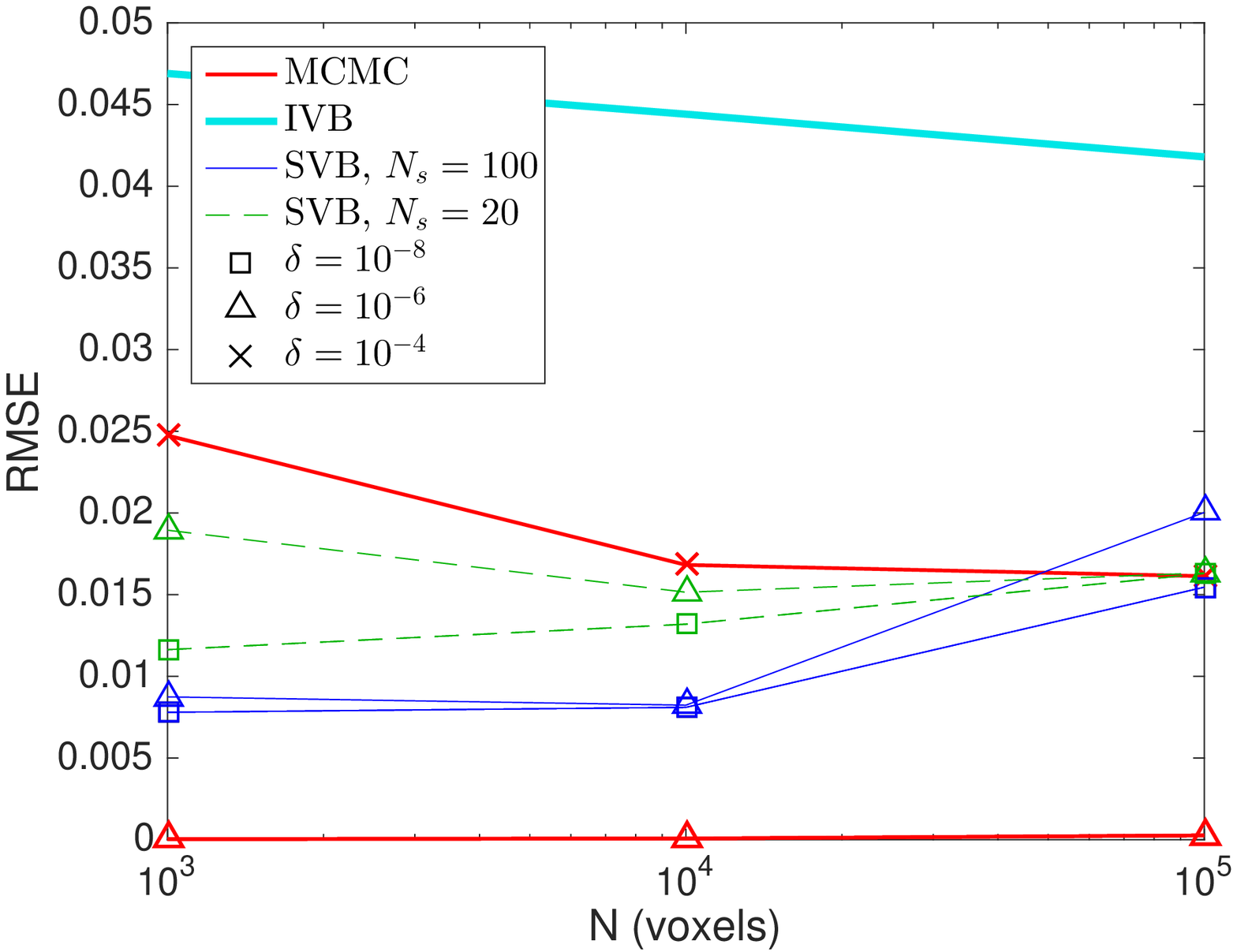}}

\caption{(a) Processing time as a function of the number of voxels for the
different algorithms and different PCG tolerance levels $\delta$.
For SVB, the number of MC samples $N_{s}$ and computing cores are
also varied. The computations were made on simulated data with $K=5$
using the 3D prior. (b) RMSE of activation coefficient posterior mean
for the different algorithms as compared to the MCMC posterior with
$\delta=10^{-8}$, based on the first $4$ regressors.\label{fig:BySize} }
\end{figure}

~\\
We simulate synthetic data from the model, with $K=5$ and $P=1$
and pick parameter values for the voxel intercept and noise standard
deviation that approximately match those of the face repetition data
described below. Several different values of $\boldsymbol{\alpha}$
were used to simulate conditions with varying informativeness, see
Appendix \ref{appendix:Data details} for details. We run the algorithms
on simulations of $N=10^{3}$, $10^{4}$ and $10^{5}$ voxels to get
an idea how they scale and compare with the number of voxels. Figure
\ref{fig:timingBySize} shows the processing time until convergence,
defined as the time until the estimated posterior mean of $\boldsymbol{\alpha}$
reaches within $1\%$ of its final value, for respective algorithm.
In Figure \ref{fig:RMSEBySize}, the accuracy of respective algorithm
is evaluated. This is based on the root-mean-squared-error (RMSE)
of the marginal posterior mean of activation coefficients $\mathbf{W}$
compared to the posterior mean from MCMC with $\delta=10^{-8}$.

Figure \ref{fig:BySize} indicates that the SVB and MCMC algorithms
scale much better with the number of voxels than IVB does, and also
provide a higher accuracy. Lowering the PCG tolerance $\delta$ from
$10^{-8}$ to $10^{-6}$ gives some speedup while seemingly sacrificing
little in accuracy. However, for $\delta=10^{-4}$ the RMSE for MCMC
increases and for SVB it becomes as high as $0.25$ with very noisy
convergence times (not shown), so this tolerance must not be set too
high. The speedup achieved by lowering $N_{s}$ seems almost linear,
but results in a lower accuracy. The speed gain in parallelizing is
small due to overhead costs, but we expect greater speed-ups for larger
data sets when each VB iteration requires more time, which is seen
for the real data in Table \ref{tab:timingByReal} below. Note that
the timing results in Figure \ref{fig:BySize} (and also in Table
\ref{tab:timingByReal}) need to be interpreted with caution since
they are based on single runs of the stochastic MCMC and SVB methods,
but these graphs provide insight about how the timing largely compares
for the different methods and settings.

\subsection{Real data}

~\\
Two real task-fMRI data sets are considered, the face repetition data
used in \citet{pennyEtAlSpatialPrior2005} and data from a visual
object recognition experiment from the OpenfMRI database \citep{Poldrack2013},
see Appendix \ref{appendix:Data details} for more details on these
data sets. 

Approximate processing times for these data are shown in Table \ref{tab:timingByReal},
both for slice-wise analysis using the 2D prior, and whole-brain analysis
using the 3D prior. The IVB method scales much worse with the number
of voxels and is hence slow in the 3D case, while MCMC and SVB are
not necessarily slower in 3D than in 2D. For whole-brain inference
with the 3D prior, SVB is generally the fastest option, especially
when run in parallel.
\begin{table}[H]
\begin{tabular}{lccccccc}
\hline 
 & \multicolumn{2}{c}{Face repetition} &  &  & \multicolumn{2}{c}{Object recognition} & \tabularnewline
 & 2D prior & 3D prior &  &  & 2D prior & 3D prior & \tabularnewline
\cline{2-3} \cline{6-7} 
IVB & 4.9 & 190 &  &  & 1.9 & 26 & \tabularnewline
MCMC & 110 & 150 &  &  & 230 & 76 & \tabularnewline
SVB & 5.3 & 11 &  &  & 22 & 20 & \tabularnewline
SVB, 4 cores & 2.8 & 3.8 &  &  & 8.7 & 7.6 & \tabularnewline
\hline 
\end{tabular}\caption{Approximate processing times (h) for the different real data sets
and algorithms using the 2D/3D prior.\label{tab:timingByReal}}
\end{table}

\begin{figure}[H]
\includegraphics[width=1\linewidth]{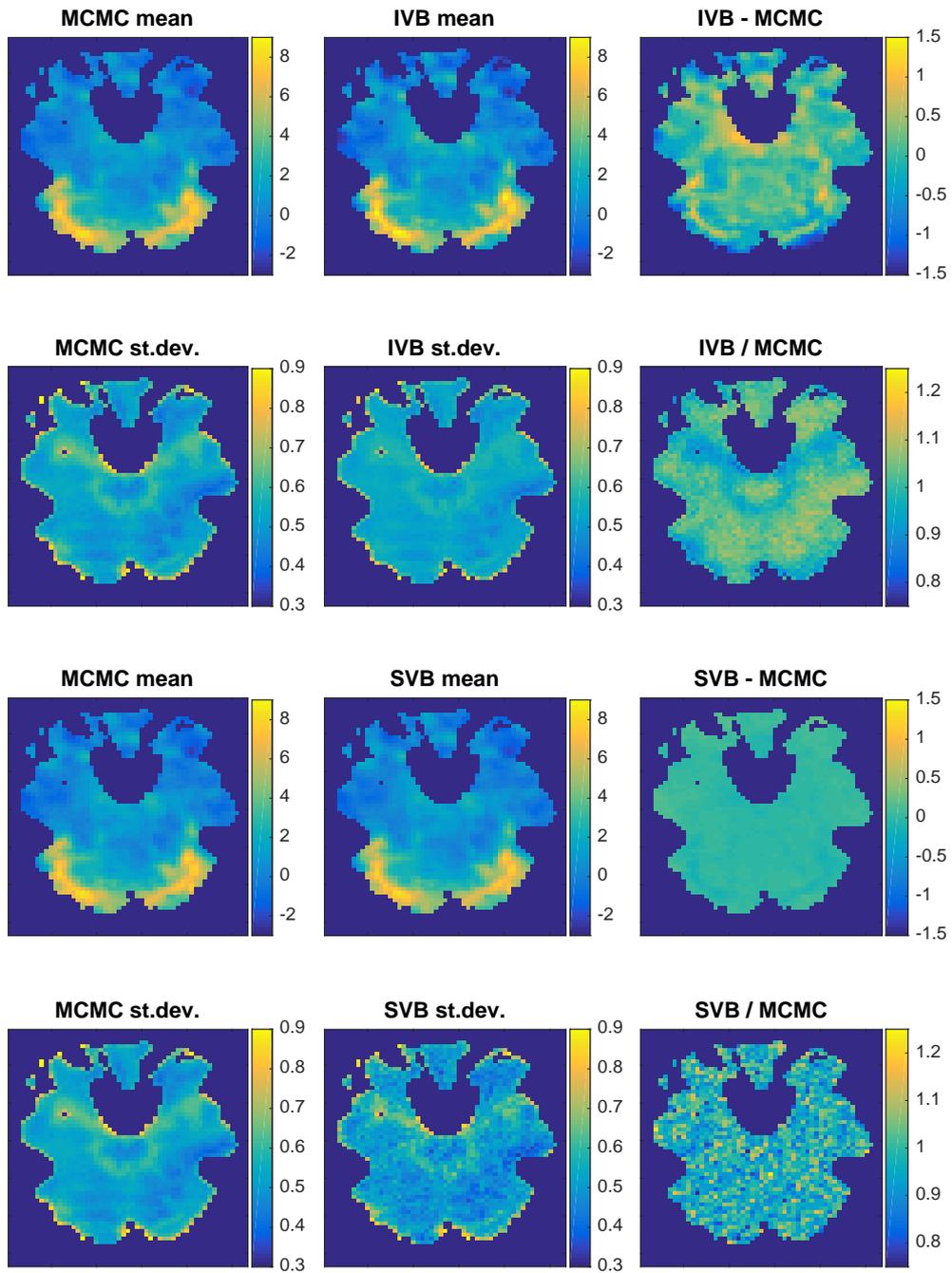}

\caption{Posterior mean (top row) and standard deviation (second row) estimated
using MCMC (left) and IVB (middle) for the contrast (mean effect of
faces) using the 2D prior. The right column shows the differences
in mean and standard deviation ratio of the estimated posteriors.
Row three and four show the corresponding results for SVB.\label{fig:MeanStdMCMCvsSPMsVB} }
\end{figure}

We first consider the face repetition data. Figure \ref{fig:MeanStdMCMCvsSPMsVB}
shows the posterior mean and standard deviation of the activation
contrast, estimated using all three methods for the 2D prior. Comparing
IVB to MCMC, we see bias both in the estimated IVB mean (the maximum
error is 1.4 across all voxels) and standard deviation (maximum error
25\%). A separate experiment with $\boldsymbol{\alpha}$ fixed to
the same value for both methods showed the well-known systematic underestimation
of standard deviations by IVB (roughly by 8\%, results not shown).
However, IVB tends to also underestimate $\boldsymbol{\alpha}$, as
shown in Figure \ref{fig:HyperPosteriors} below, leading to less
shrinkage and VB errors in standard deviation that go in both directions.
Comparing SVB to MCMC, we see that the estimated SVB mean is much
more correct than the IVB (maximum error 0.2). However, the SVB standard
deviation estimates are quite inaccurate (maximum error 26\%), but
this is mainly due to the noisy covariance estimates discussed in
Section \ref{subsec:Spatial-VB}.

\begin{figure}[H]
\includegraphics[width=1\linewidth]{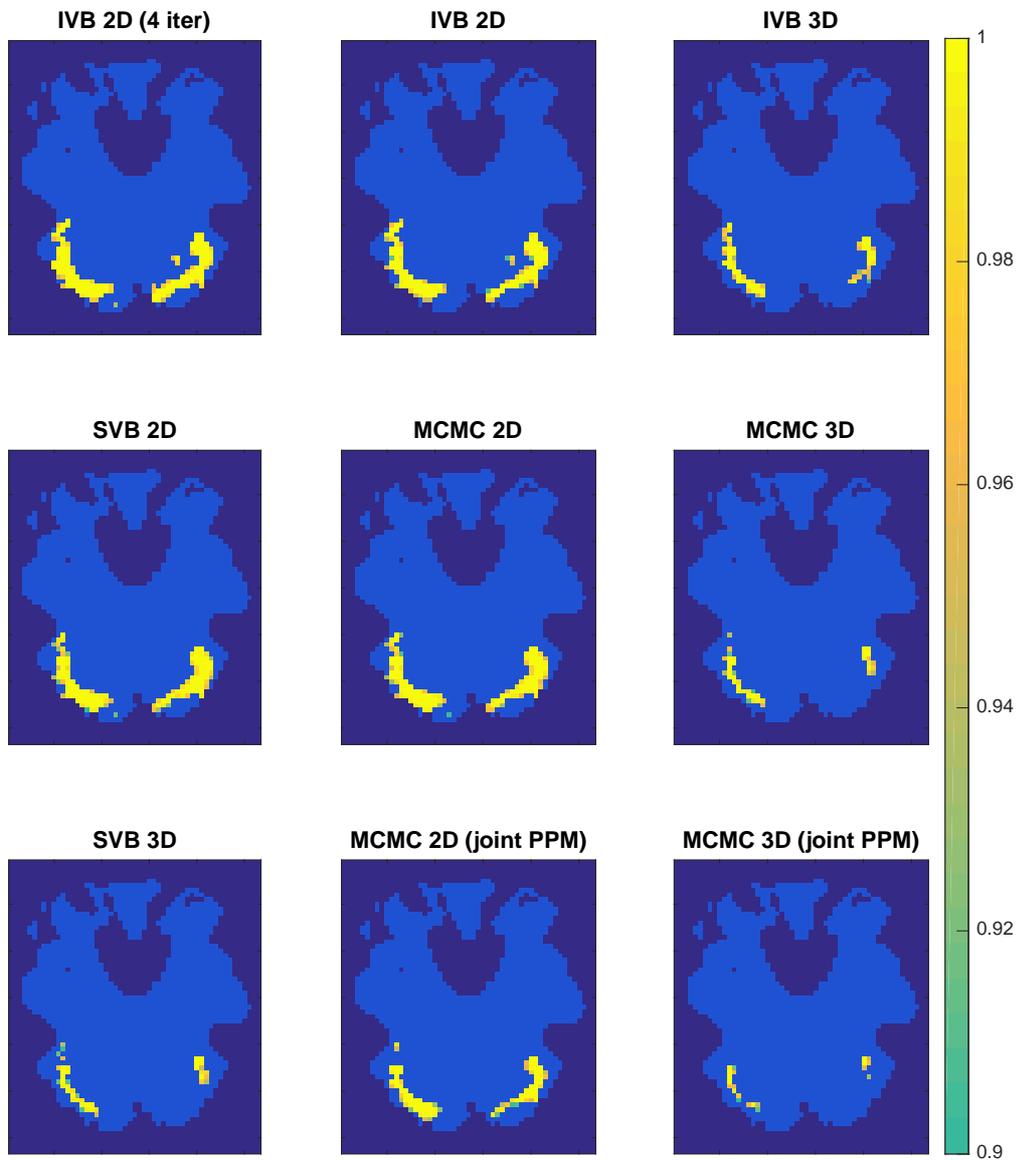}

\caption{PPMs estimated for the face repetition data with different methods
and 2D or 3D spatial priors. The PPMs show probabilities of the contrast
(mean effect of faces) exceeding 1\% of the global mean signal, thresholded
at 0.9.\label{fig:faceRepPPMs} }
\end{figure}

Computed marginal PPMs for various settings, thresholded at 0.9, are
shown in Figure \ref{fig:faceRepPPMs}. The first row shows estimates
for IVB after 4 VB iterations for the 2D prior (which is the SPM12
default), and after convergence for both the 2D and 3D prior. In the
second and third row one can see the corresponding PPMs estimated
using MCMC and SVB, and also the MCMC based joint PPMs which were
computed using the R package \texttt{excursions}. The 2D joint PPM
is computed based only on voxels within this slice while the 3D joint
PPM is computed based on all voxels in the brain, so these maps are
not directly comparable. The differences between the non-converged
and converged SPM results seem rather small for this data set. Overall,
the PPMs from IVB agree quite well with the PPMs from MCMC; an exception
is a small cluster of activity in IVB which is lacking in the activation
map from MCMC. The MCMC and SVB PPMs are hard to distinguish. Larger
differences are found when comparing results based on the 2D vs. 3D
prior and when comparing marginal vs. joint PPMs. 
\begin{figure}[H]
\includegraphics[width=1\linewidth]{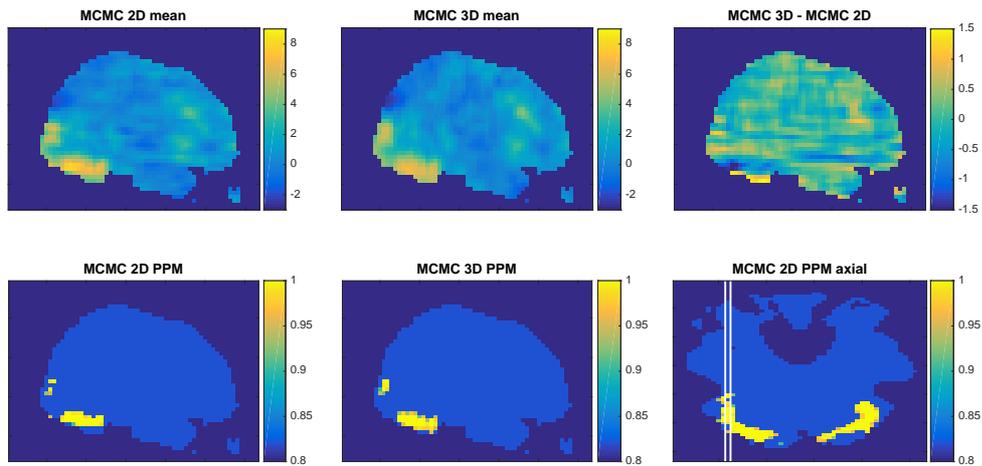}

\caption{Posterior mean (top row) and marginal PPMs (bottom row) for the face
repetition data for the MCMC method when using the 2D and 3D spatial
priors for the sagittal slice indicated in the bottom right figure.
The PPMs show probabilities of the contrast (mean effect of faces)
exceeding 1\% of the global mean signal, thresholded at 0.9.\label{fig:faceRepMeanPPMMCMC} }
\end{figure}
The smaller activity regions from the 3D prior compared to the 2D
prior is due to shrinkage towards a larger number of non-active voxels
nearby in the $z$-dimension, which is depicted in Figure \ref{fig:faceRepMeanPPMMCMC}
that compares the posterior mean and PPM for the two priors in a sagittal
slice. It is clear that the posterior mean is generally lower in the
active regions when using the 3D prior, which can only be explained
with the assumed dependence with non-active voxels in the $z$-dimension.
For this particular slice, the effect is nevertheless strong enough
for these voxels to be classified as active in the PPMs, but for other
slices the higher smoothness can bring posterior probabilities below
the $0.9$-threshold, for voxels classified as active when using the
2D prior. At the same time, for the 2D prior we observe discontinuous
effects between slices, for example in the inferior part of the largest
active blob, while the 3D prior lends strength to the voxels below,
classifying them as active. It should be noted that the 2D and 3D
prior lead to rather different models (the 2D prior implies a different
covariance structure and has many more parameters and is therefore
more flexible), so the results are not directly comparable and greatly
data dependent. To decide which of these priors (or perhaps a more
flexible 3D prior or parcel based method) is best is a model selection
problem, which we see as important future work.
\begin{figure}[H]
\includegraphics[width=1\linewidth]{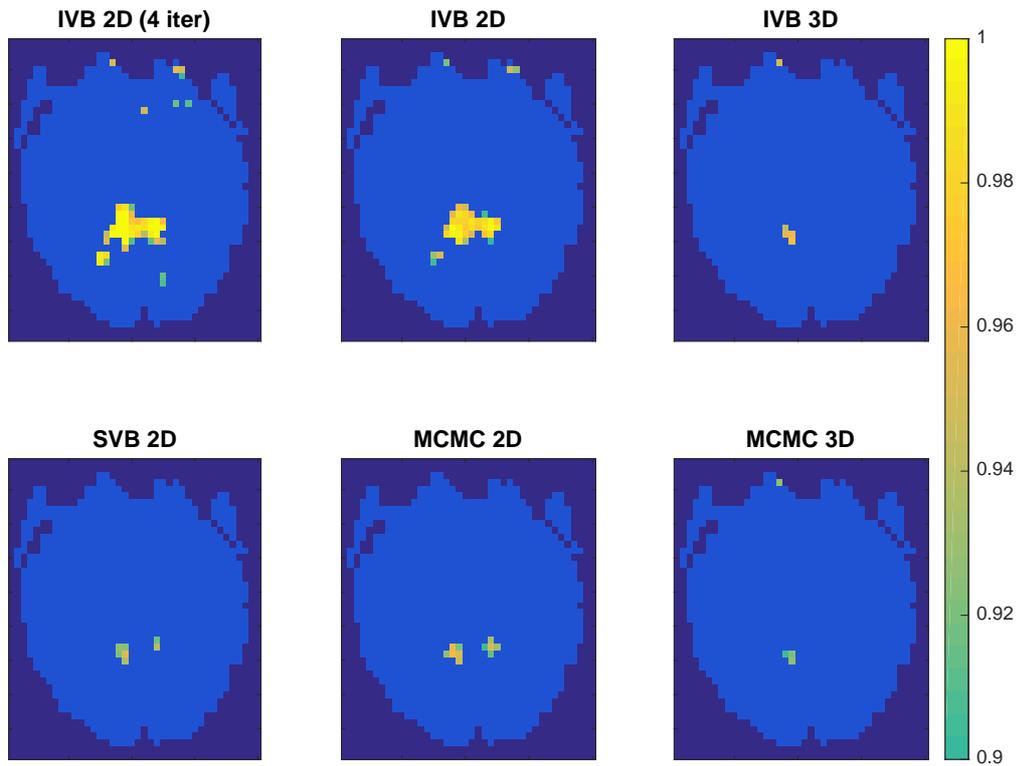}

\caption{Marginal PPMs estimated for the object recognition data with different
methods and 2D or 3D spatial priors. The PPMs show probabilities of
the contrast (houses vs. faces) exceeding 0.5\% of the global mean
signal, thresholded at 0.9.\label{fig:openfMRIPPMs} }
\end{figure}

Although we observe some errors in the posterior mean and standard
deviation for IVB as compared to MCMC, this error is not big enough
to make much impact on the PPMs for the face repetition data set.
Our second dataset, the object recognition data, shows a situation
where this error can in fact be severe also for the PPMs. Figure \ref{fig:openfMRIPPMs}
shows computed PPMs for one slice of the object recognition data,
for some different methods and using the 2D/3D prior. It is clear
that the independence assumption in IVB can lead to severely distorted
activation maps, and that SVB is a much more accurate approximation
for this dataset. 

Much of the differences in brain maps between IVB and the other methods
can be attributed to the underestimation of hyperparameters. Figure
\ref{fig:HyperPosteriors} shows the estimated posteriors of the spatial
hyperparameters $\boldsymbol{\alpha}$ and $\boldsymbol{\beta}$ for
the main regressors and AR coefficient for the different methods and
data sets. When data are informative, as for the intercept, the VB
methods generally approximate the hyperparameter posteriors well,
but as data become less informative (this is clearly seen for the
simulated data) the approximate posteriors from IVB underestimate
both the location and dispersion of the posterior. The underestimation
of posterior dispersion is a well known issue with VB quite generally
\citep{Bishop2006}, but the inability of IVB to correctly approximate
the posterior location is unusual. The answer comes from the assumption
of posterior independent voxels. Since the noise is assumed to be
spatially independent, the posterior dependence between voxel activations
comes solely from the spatial prior. IVB therefore pushes the $\alpha_{k}$:s
to lower values in an attempt to reduce the influence of the prior.
Indeed, SVB is much better at finding the correct location, but, like
most VB methods, tends to underestimate the dispersion. 
\begin{figure}[H]
\subfloat[\label{fig:HyperPosteriorsSimulated}]{\includegraphics[width=0.5\linewidth]{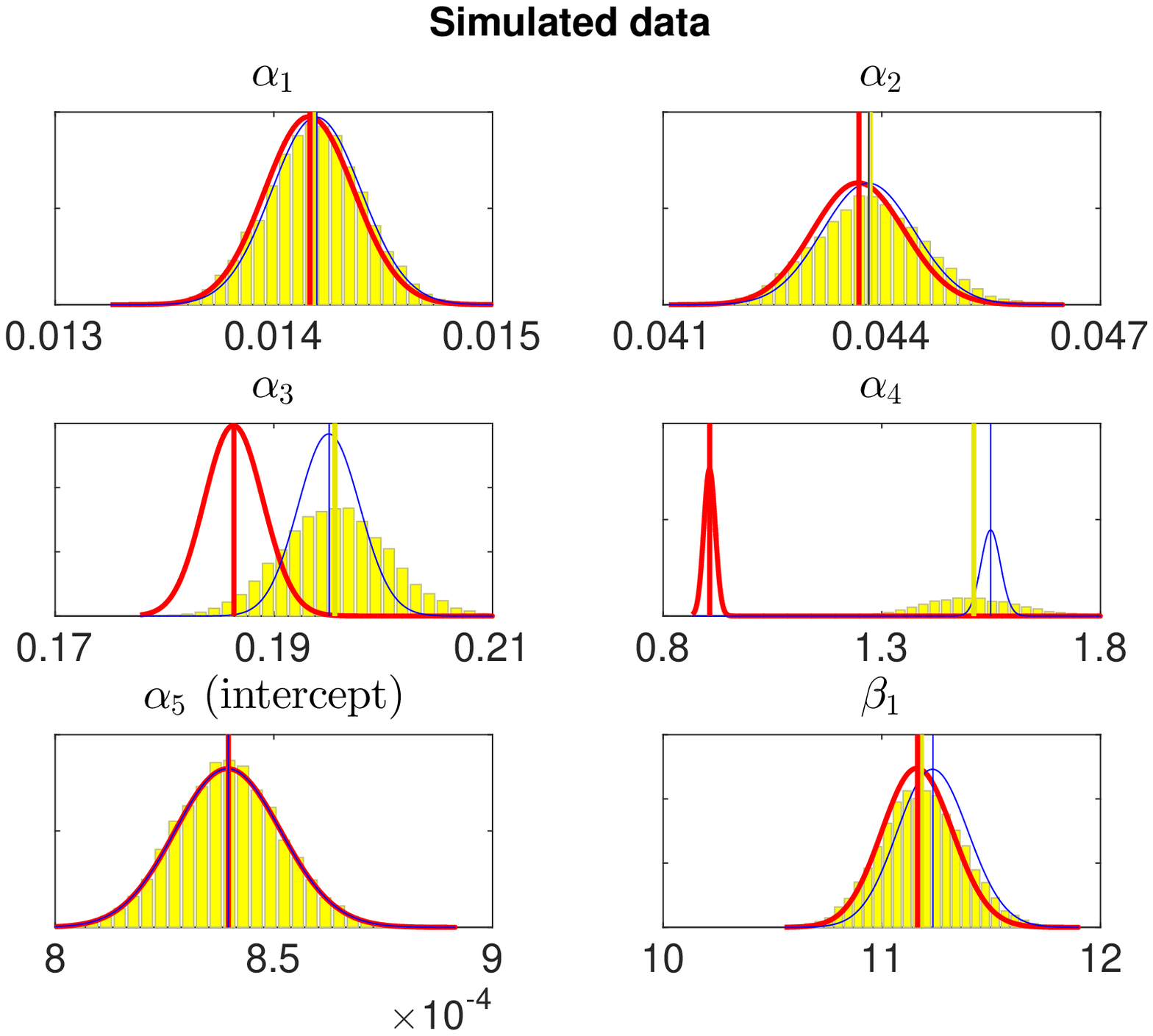}}\subfloat[\label{fig:HyperPosteriorsFaceRep}]{\includegraphics[width=0.5\linewidth]{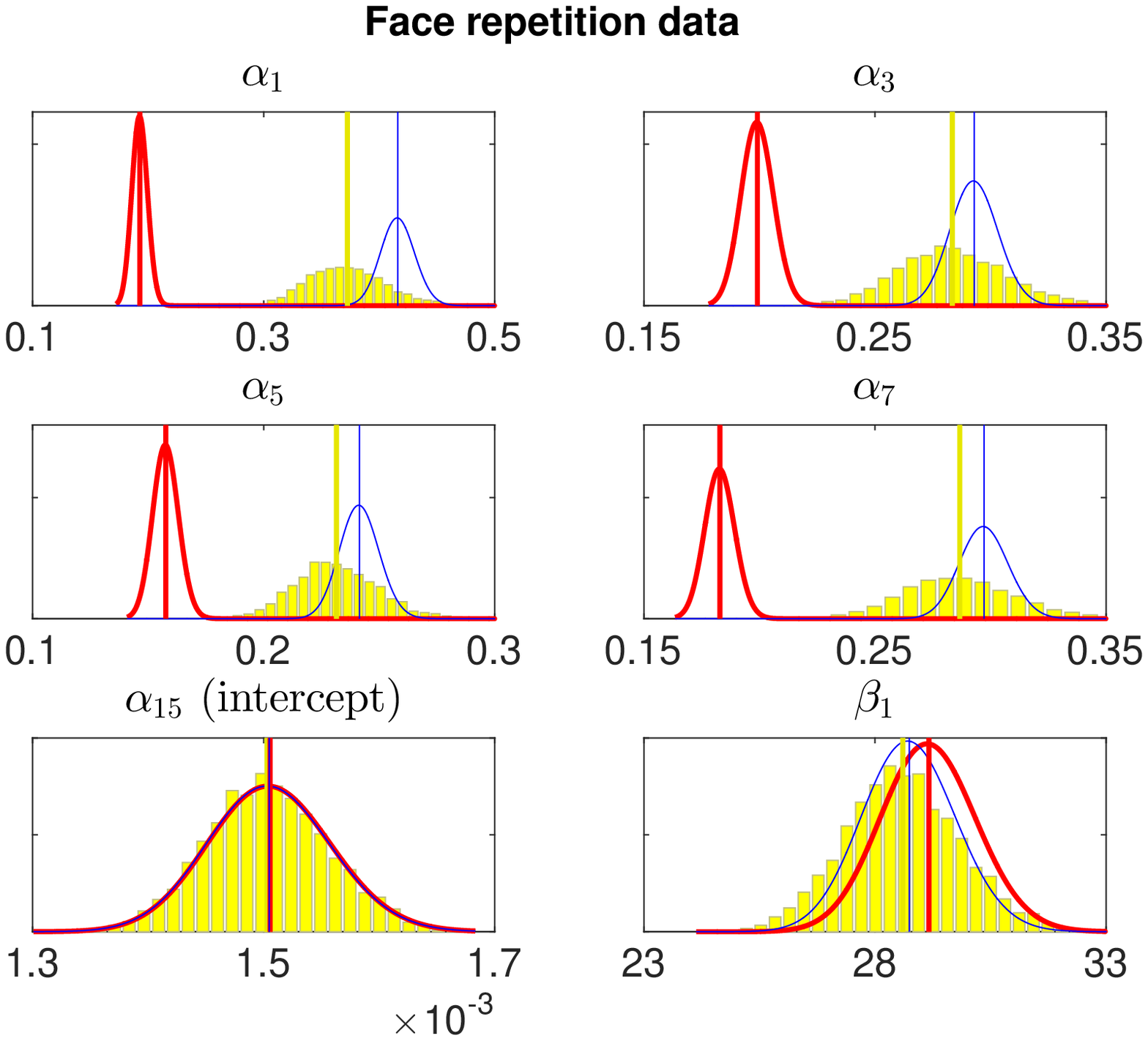}}

\subfloat[\label{fig:HyperPosteriorsObjRec}]{\includegraphics[width=1\linewidth]{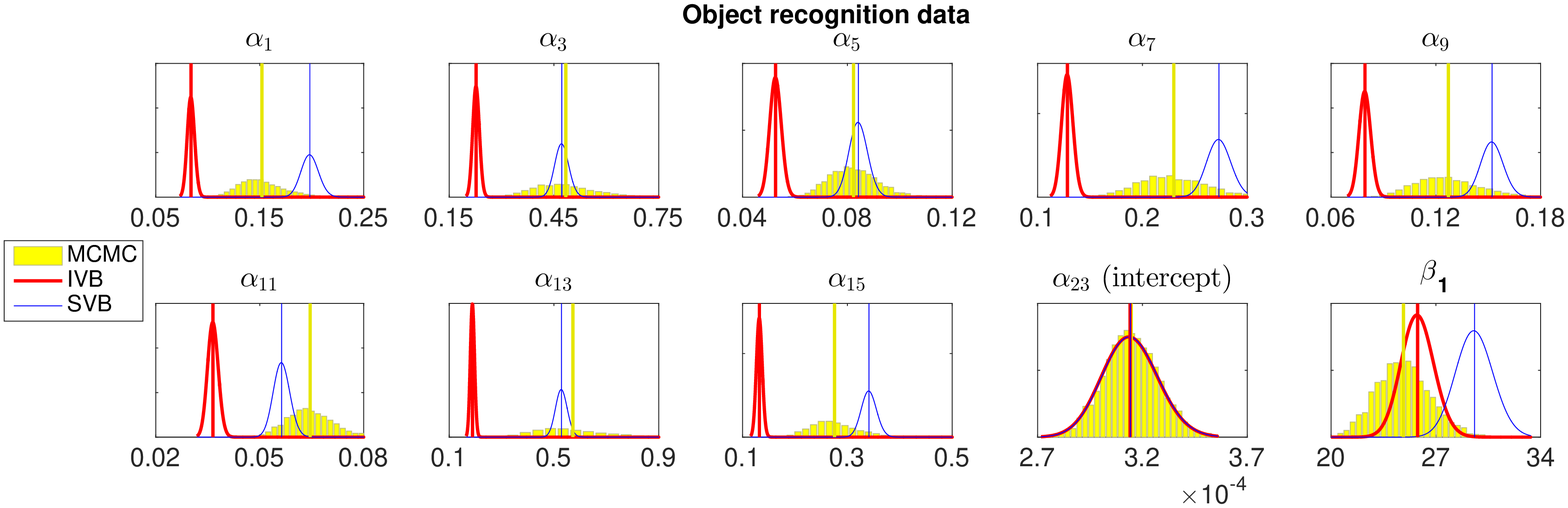}}

\caption{Estimated posteriors for the spatial hyperparameters for the main
regressors and the first AR coefficient for the simulated (a), face
repetition (b) and object recognition (c) data by MCMC (histogram),
IVB (red) and SVB (blue). \label{fig:HyperPosteriors} }
\end{figure}

\section{Discussion and future work}

The PCG based GMRF sampling provides a fast way to perform exact inference
in high-dimensional spatial models such as the one considered here
for task-fMRI. The presented results show that our methods scale better
than SPM's IVB, which is simultaneously shown to generate erroneous
results for certain data.

\subsubsection*{VB approximation error}

The IVB error comes from two sources. Firstly, factorized VB is known
to underestimate posterior variances \citep{Bishop2006} and we have
seen that fixing $\boldsymbol{\alpha}$ to the same value in both
SPM's VB and the MCMC algorithms results in underestimated $\mathbf{W}$
posterior standard deviation for IVB for the face repetition data.
The same type of posterior variance underestimation can be seen for
the SVB hyperparameters in Figure \ref{fig:HyperPosteriors}, a behavior
that is theoretically motivated in \citet{Solis-Trapala2009} (Appendix
A). Secondly, we have seen that IVB tends to underestimate also the
mean of $\boldsymbol{\alpha}$ for many regressors, resulting in the
wrong level of smoothing/shrinkage. 

The SVB method seems to approximate the exact $\mathbf{W}$ posterior
well in most cases, but the underestimation of hyperparameter variance
is occasionally quite large (sometimes also the mean is slightly wrong)
which could motivate dropping the VB assumptions entirely and instead
optimizing these parameters and perhaps using a Gaussian approximation
for the uncertainty.

\subsubsection*{Computational improvements and model extensions}

There is room for further improvement of the SVB method, by investigation
of the sensitivity to settings like the PCG tolerance $\delta$ and
the number of MC samples $N_{s}$ on a larger number of data sets.
Different pre-conditioners could be used, for example the robust incomplete
Cholesky \citep{Ajiz1984}, and \citet{Bolin2014b} discuss other
approximations of traces like the one in equation (\ref{eq:quad-form-expectation}).
The evaluation should be in relation to the output of interest, for
example PPMs, to find the optimal balance between accuracy and processing
time. In addition, a better online criterion for convergence would
be beneficial. In future work we will also investigate if graphics
processing units (GPUs) can be used to reduce the processing time
further \citep{eklund_gpu,broccoli}.

Even though the MCMC algorithm is probably not fast enough for the
everyday practitioner to run on whole-brain data sets with many conditions,
it is orders of magnitude faster than what would be the case without
PCG sampling, in which case it would be impractical to run at all.
It fulfills an important purpose as the ground truth when evaluating
approximate methods such as the VB methods in this paper, and could
practically be used on sub-volumes of interest. 

For the algorithms presented in this paper we have seen that the convergence
rate is determined by the spatial hyperparameters for the least informative
regressors. Hence, a small model change that would increase the convergence
rates could be to drop the spatial prior on regressors that are very
non-informative (for example motion regressors, as previously suggested
by \citet{Groves2009}).

In this work, we have for brevity only considered the spatial UGL
prior and focused on exact inference using this approximate model.
For example, the UGL prior is stationary, isotropic and cannot separate
shrinkage and smoothing. As mentioned in the introduction, many alternative,
less approximate, priors have been proposed which would be interesting
to adopt to the PCG sampling framework for 3D inference. Many of these
would be straightforward to implement, for example the anatomically
motivated tissue-type AR-priors in \citet{Penny2007}, while others
would require solving additional computational issues. A particularly
interesting alternative would be the Matérn kernel which is a standard
choice in spatial statistics. The Matérn kernel defines a Gaussian
field (GF) that in general is not Markov, however, \citet{Lindgren2011}
give an explicit link between GFs and GMRFs for the Matérn class such
that it can be used in a sparsity exploiting manner, and they also
provide a possible non-stationary Matérn model. Other parallel model
improvements would be to include a spatial model also for the noise
precision $\boldsymbol{\lambda}$, to add a spatial prior also for
the probability of activation and to include spatial dependence in
the likelihood, which is motivated by \citet{Kriegeskorte2008} who,
using a phantom, demonstrate that noise from echoplanar imaging is
naturally spatial.

While the voxels are normally equally sized in the $x$ and $y$ direction
of each slice, this is not necessarily the case relative to the $z$
direction and one might worry that this leads to more anisotropic
data that cannot be accounted for by the UGL prior. A simple solution
to this is to resample the voxels to have the same size during the
pre-processing, which is in fact what we did for the face repetition
data, but not for the object recognition data. The same solution is
possible for the second data set, but another straightforward solution
would be to replace the UGL prior with a weighted graph-Laplacian
(WGL) prior, that is to redefine $\mathbf{D}_{w}=\mathbf{G}_{w}^{\prime}\mathbf{C}\mathbf{G}_{w}$
instead of $\mathbf{D}_{w}=\mathbf{G}_{w}^{\prime}\mathbf{G}_{w}$
for a diagonal weight matrix $\mathbf{C}$ that can be chosen based
on the Euclidian distances between neighboring voxels as in \citet{Harrison2008}.
This solution would also be simply adapted to the PCG framework by
just multiplying row $i$ in $\mathbf{G}_{w}$ with $\sqrt{C_{ii}}$
when using Algorithm \ref{Alg:PCG sampling}. Another solution would
be change the prior precision matrix in equation (\ref{eq:Wk prior})
from $\alpha_{k}\mathbf{G}_{w}^{\prime}\mathbf{G}_{w}$ to $\mathbf{G}_{w}^{\prime}\mathbf{C}_{k}\mathbf{G}_{w}$
with
\[
\mathbf{C}_{k}=\left[\begin{array}{ccc}
\alpha_{x,k}\mathbf{I}\\
 & \alpha_{y,k}\mathbf{I}\\
 &  & \alpha_{z,k}\mathbf{I}
\end{array}\right],
\]
with $\alpha_{x,k}$, $\alpha_{y,k}$ and $\alpha_{z,k}$ being random
parameters (to be inferred) that applies to neighboring pairs of voxels
in the $x$, $y$ and $z$ direction respectively. Adopting such a
model to the MCMC or SVB frameworks would however be difficult for
computational reasons. 

For inference using this or other more advanced priors than the UGL,
Gibbs sampling is often not feasible. In these situations, or when
the mixing of the Gibbs sampling chain is dissatisfactory, one can
instead attempt to perform MCMC using Metropolis-Hastings (MH) steps
or Hamiltonian Monte Carlo (HMC) \citep{Duane1987,Neal2011}. These
methods require the computation of acceptance probabilities based
on the joint posterior ratio, as demonstrated for our model in Appendix
\ref{Appendix full conditionals derivations}. These become computable
in practice because the hyperparameters $\alpha_{k}$ can be factored
out of the determinant of the prior precision matrix as in equation
(\ref{eq:UnnormalisedWPrior}), but for a more advanced prior the
computation of this determinant would be problematic. In a recent
work, independent to ours, \citet{Teng2016} use HMC to do inference
in the same model as the one in our article. They also make comparisons
to SPM's VB for the face repetition data using the 3D prior, and obtain
similar results. Their reported computing times seem fast, but since
they are using a much smaller model $\left(K=5\right)$ and a different
implementation (C++), another number of samples, etc., their computing
times cannot be directly compared to ours. Nevertheless, HMC seems
like a competitive method to the proposed PCG based Gibbs sampling
method for MCMC in these models. In a different, recent and independent
article, \citet{Rad} use a similar PCG based Gibbs sampling method
to ours in a similar model for other kinds of neuroscientific data.
However, they use a different spatial edge-preserving Laplace prior,
which would be interesting to apply also to fMRI data within this
framework.

In order to be able to properly motivate any of the above mentioned
model improvements, however, a model selection criterion is required.
\citet{Penny2007} used the model evidence lower bound, which is a
good alternative, but requires the computation of determinants of
precision matrices of size $KN\times KN$, which is infeasible in
the 3D case. Good approximations of such determinants would therefore
be an eligible direction for future research. Model selection criteria
based on cross validation or the marginal likelihood could also be
explored.

\subsubsection*{Multiple comparisons}

There is currently no consensus on how to control for multiple comparisons
in Bayesian spatial models for fMRI data. We showed how joint PPMs
based on excursion sets can be computed for the MCMC method, but the
large data size is currently preventing us from computing the same
for the SVB method. The joint PPMs control the family-wise error rate
given the spatial model and a threshold, but a separate objective
could be to instead control the false discovery rate (FDR) for which
a unified framework within large-scale spatial models is provided
by \citet{Sun2015}. An interesting area of future research would
thus be to adopt both these approaches to the 3D spatial modeling
of fMRI data when using both the MCMC and SVB method.

\subsubsection*{Group analysis}

Another important area of future work is to extend this single subject
analysis to the group level. The simplest way to do this is to consider
the posterior mean maps from the single subject analysis as spatially
processed and use these as input to a voxel-wise Bayesian regression,
which is basically what is done in SPM's Bayesian second level analysis.
This approach has several drawbacks, one being that mis-registration
between subjects causes activation to be located in different voxels,
and can therefore be missed when averaging across subjects. The classic
GLM approach ``handles'' this by using smoothing as a pre-processing
step. A more elegant, Bayesian solution is presented in \citet{Xu2009}
which explicitly models population level activation centers. A second
drawback is that such a procedure discards the posterior uncertainty
from the subject level analyses. The most natural way to do the group
level analysis would instead be using a Bayesian hierarchical model
with a common group level activation working as a latent prior for
each subject, and to also model the spatial hyperparameters at the
group level. However, estimating such a model would intuitively be
very demanding in terms of memory and speed, and an approximate idea
is to instead target the mean of the subject-level activations, as
in \citet{Yue2014}. 

\subsection{In conclusion}

~\\
A fast and practical MCMC scheme for exact whole-brain spatial inference
in task-fMRI is suggested and implemented. Also, a non-factorizing
VB method is developed and shown to give practically the same results,
but in shorter time. The methods are compared to the popular factorizing
VB method in SPM and shown to scale better with problem size. The
comparison with the exact MCMC estimates gives evidence that SPM's
VB can produce false activity estimates in some settings.

\section*{Acknowledgements}

We thank Will Penny and three anonymous reviewers for helpful comments.
This work was funded by Swedish Research Council (Vetenskapsrådet)
grant no 2013-5229. David Bolin was also supported by the Knut and
Alice Wallenberg foundation.\bibliographystyle{apalike}
\bibliography{library,manually}

\begin{thebibliography}{}

\bibitem[SPM, 2002]{SPMsoftware}
 (2002).
\newblock {\em SPM}.
\newblock Wellcome Department of Imaging Neuroscience, Available at
  http://www.fil.ion.ucl.ac.uk/spm/software.

\bibitem[Ajiz and Jennings, 1984]{Ajiz1984}
Ajiz, M.~A. and Jennings, A. (1984).
\newblock {A robust incomplete Cholesky-conjugate gradient algorithm}.
\newblock {\em Int. J. Numer. Meth. Engng.}, 20:949--966.

\bibitem[Amestoy et~al., 1996]{Amestoy1996}
Amestoy, P.~R., Davis, T.~A., and Duff, I.~S. (1996).
\newblock {An Approximate Minimum Degree Ordering Algorithm}.
\newblock {\em SIAM. J. Matrix Anal. {\&} Appl.}, 17(4):886--905.

\bibitem[Barrett et~al., 1994]{barrett1994templates}
Barrett, R., Berry, M.~W., Chan, T.~F., Demmel, J., Donato, J., Dongarra, J.,
  Eijkhout, V., Pozo, R., Romine, C., and Van~der Vorst, H. (1994).
\newblock {\em Templates for the solution of linear systems: building blocks
  for iterative methods}, volume~43.
\newblock Siam.

\bibitem[Besag, 1974]{Besag1974}
Besag, J. (1974).
\newblock {Spatial Interaction and the Statistical Analysis of Lattice
  Systems}.
\newblock {\em Journal of the Royal Statistical Society. Series B
  (Methodological)}, 36(2):192--236.

\bibitem[Bishop, 2006]{Bishop2006}
Bishop, C.~M. (2006).
\newblock {\em {Pattern Recognition and Machine Learning}}.
\newblock Springer.

\bibitem[Bolin and Lindgren, 2015]{Bolin2014a}
Bolin, D. and Lindgren, F. (2015).
\newblock {Excursion and contour uncertainty regions for latent Gaussian
  models}.
\newblock {\em Journal of the Royal Statistical Society: Series B (Statistical
  Methodology)}, 77(1):85--106.

\bibitem[Bolin et~al., 2014]{Bolin2014b}
Bolin, D., Wallin, J., and Lindgren, F. (2014).
\newblock {Multivariate latent Gaussian random field mixture models}.
\newblock {\em Preprint Dept. Math. Sci., Chalmers University of Technology and
  G{\"{o}}teborg University 2014:1}.

\bibitem[Chaari et~al., 2013]{Chaari2013}
Chaari, L., Vincent, T., Forbes, F., Dojat, M., and Ciuciu, P. (2013).
\newblock {Fast Joint Detection-Estimation of Evoked Brain Activity in
  Event-Related fMRI Using a Variational Approach Lotfi}.
\newblock {\em IEEE Transactions on Medical Imaging}, 32(5):821--837.

\bibitem[Chan and Ledolter, 1995]{Chan1995}
Chan, K.~S. and Ledolter, J. (1995).
\newblock {Monte Carlo EM Estimation for Time Series Models Involving Counts}.
\newblock {\em Journal of the American Statistical association},
  90(429):242--252.

\bibitem[Delyon et~al., 1999]{Delyon1999}
Delyon, B., Lavielle, M., and Moulines, E. (1999).
\newblock {Convergence of a stochastic approximation version of the EM
  algorithm}.
\newblock {\em Annals of Statistics}, 27(1):94--128.

\bibitem[Duane et~al., 1987]{Duane1987}
Duane, S., Kennedy, A.~D., Pendleton, B.~J., and Roweth, D. (1987).
\newblock {Hybrid Monte Carlo}.
\newblock {\em Physics letters B}, 195(2):216--222.

\bibitem[Eklund et~al., 2013]{eklund_gpu}
Eklund, A., Dufort, P., Forsberg, D., and LaConte, S. (2013).
\newblock Medical image processing on the {GPU} - {Past}, present and future.
\newblock {\em Medical Image Analysis}, 17:1073--1094.

\bibitem[Eklund et~al., 2014]{broccoli}
Eklund, A., Dufort, P., Villani, M., and LaConte, S. (2014).
\newblock {BROCCOLI: Software for fast fMRI analysis on many-core CPUs and
  GPUs}.
\newblock {\em Frontiers in Neuroinformatics}, 8:24.

\bibitem[Eklund et~al., 2016]{Eklund2016}
Eklund, A., Nichols, T.~E., and Knutsson, H. (2016).
\newblock {Cluster failure: why fMRI inferences for spatial extent have
  inflated false positive rates}.
\newblock {\em Proceedings of the National Academy of Sciences},
  113(28):7900--7905.

\bibitem[Friston et~al., 1995]{Friston1995a}
Friston, K.~J., Holmes, a.~P., Worsley, K.~J., Poline, J.-P., Frith, C.~D., and
  Frackowiak, R. S.~J. (1995).
\newblock {Statistical parametric maps in functional imaging: A general linear
  approach}.
\newblock {\em Human Brain Mapping}, 2(4):189--210.

\bibitem[Friston and Penny, 2003]{Friston2003a}
Friston, K.~J. and Penny, W. (2003).
\newblock {Posterior probability maps and SPMs}.
\newblock {\em NeuroImage}, 19(3):1240--1249.

\bibitem[Groves et~al., 2009]{Groves2009}
Groves, A.~R., Chappell, M.~A., and Woolrich, M.~W. (2009).
\newblock {Combined spatial and non-spatial prior for inference on MRI
  time-series}.
\newblock {\em NeuroImage}, 45(3):795--809.

\bibitem[Gunawan et~al., 2016]{Gunawan2016}
Gunawan, D., Tran, M., and Kohn, R. (2016).
\newblock {Fast Inference for Intractable Likelihood Problems using Variational
  Bayes}.
\newblock {\em preprint: http://hdl.handle.net/2123/14594}.

\bibitem[Hanson et~al., 2004]{Hanson2004a}
Hanson, S.~J., Matsuka, T., and Haxby, J.~V. (2004).
\newblock {Combinatorial codes in ventral temporal lobe for object recognition:
  Haxby (2001) revisited: is there a "face" area?}
\newblock {\em NeuroImage}, 23(1):156--166.

\bibitem[Harrison and Green, 2010]{Harrison2010}
Harrison, L.~M. and Green, G. G.~R. (2010).
\newblock {A Bayesian spatiotemporal model for very large data sets}.
\newblock {\em NeuroImage}, 50(3):1126--1141.

\bibitem[Harrison et~al., 2008a]{Harrison2008}
Harrison, L.~M., Penny, W., Daunizeau, J., and Friston, K.~J. (2008a).
\newblock {Diffusion-based spatial priors for functional magnetic resonance
  images}.
\newblock {\em NeuroImage}, 41(2):408--423.

\bibitem[Harrison et~al., 2008b]{Harrison2008a}
Harrison, L.~M., Penny, W., Flandin, G., Ruff, C.~C., Weiskopf, N., and
  Friston, K.~J. (2008b).
\newblock {Graph-partitioned spatial priors for functional magnetic resonance
  images}.
\newblock {\em NeuroImage}, 43(4):694--707.

\bibitem[Haxby et~al., 2001]{Haxby2001a}
Haxby, J.~V., Gobbini, M.~I., Furey, M.~L., Ishai, A., Schouten, J.~L., and
  Pietrini, P. (2001).
\newblock {Distributed and Overlapping Representations of Faces and Objects in
  Ventral Temporal Cortex}.
\newblock {\em Science}, 293(5539):2425--2430.

\bibitem[Henson et~al., 2002]{Henson2002}
Henson, R., Shallice, T., Gorno-Tempini, M.~L., and Dolan, R. (2002).
\newblock {Face repetition effects in implicit and explicit memory tests as
  measured by fMRI}.
\newblock {\em Cerebral Cortex}, 12:178--186.

\bibitem[Kingma and Welling, 2014]{Kingma2013}
Kingma, D.~P. and Welling, M. (2014).
\newblock {Auto-Encoding Variational Bayes}.
\newblock {\em arXiv:1312.6114v10}.

\bibitem[Kriegeskorte et~al., 2008]{Kriegeskorte2008}
Kriegeskorte, N., Bodurka, J., and Bandettini, P. (2008).
\newblock {Artifactual time-course correlations in echo-planar fMRI with
  implications for studies of brain function}.
\newblock {\em International Journal of Imaging Systems and Technology},
  18(5-6):345--349.

\bibitem[Lindgren et~al., 2011]{Lindgren2011}
Lindgren, F., Rue, H., and Lindstr{\"{o}}m, J. (2011).
\newblock {An explicit link between Gaussian fields and Gaussian Markov random
  fields: The SPDE approach}.
\newblock {\em Journal of the Royal Statistical Society Series B},
  73(4):423--498.

\bibitem[Manteuffel, 1980]{Manteuffel1980}
Manteuffel, T.~a. (1980).
\newblock {An incomplete factorization technique for positive definite linear
  systems}.
\newblock {\em Mathematics of Computation}, 34(150):473--473.

\bibitem[Musgrove et~al., 2016]{Musgrove2015}
Musgrove, D.~R., Hughes, J., and Eberly, L.~E. (2016).
\newblock {Fast, fully Bayesian spatiotemporal inference for fMRI data}.
\newblock {\em Biostatistics}, 17(2):291--303.

\bibitem[Neal, 2011]{Neal2011}
Neal, R.~M. (2011).
\newblock {MCMC using Hamiltonian dynamics}.
\newblock {\em Handbook of Markov Chain Monte Carlo}, pages 113--162.

\bibitem[O'Toole et~al., 2005]{OToole2005a}
O'Toole, A.~J., Jiang, F., Abdi, H., and Haxby, J.~V. (2005).
\newblock {Partially Distributed Representations of Objects and Faces in
  Ventral Temporal Cortex}.
\newblock {\em Journal of Cognitive Neuroscience}, 17(4):580--590.

\bibitem[Papandreou and Yuille, 2010]{Papandreou2010}
Papandreou, G. and Yuille, A. (2010).
\newblock {Gaussian sampling by local perturbations}.
\newblock {\em Advances in Neural Information Processing Systems 23},
  90(8):1858--1866.

\bibitem[Papandreou and Yuille, 2011]{Papandreou2011}
Papandreou, G. and Yuille, A.~L. (2011).
\newblock {Efficient variational inference in large-scale Bayesian compressed
  sensing}.
\newblock {\em 2011 IEEE International Conference on Computer Vision Workshops
  (ICCV Workshops)}, pages 1332--1339.

\bibitem[Penny and Flandin, 2005]{Penny2005a}
Penny, W. and Flandin, G. (2005).
\newblock {Bayesian analysis of fMRI data with spatial priors}.
\newblock {\em In Proceedings of the Joint Statistical Meeting (JSM). American
  Statistical Association.}

\bibitem[Penny et~al., 2007]{Penny2007}
Penny, W., Flandin, G., and Trujillo-Barreto, N. (2007).
\newblock {Bayesian comparison of spatially regularised general linear models}.
\newblock {\em Human Brain Mapping}, 28(4):275--293.

\bibitem[Penny et~al., 2003]{Penny2003}
Penny, W., Kiebel, S., and Friston, K. (2003).
\newblock {Variational Bayesian inference for fMRI time series}.
\newblock {\em NeuroImage}, 19(3):727--741.

\bibitem[Penny et~al., 2005a]{Penny2005}
Penny, W.~D., Trujillo-Bareto, N., and Flandin, G. (2005a).
\newblock {Bayesian analysis of single-subject fMRI data: SPM implementation}.
\newblock {\em Technical report, Wellcome Department of Imaging Neuroscience.}

\bibitem[Penny et~al., 2005b]{pennyEtAlSpatialPrior2005}
Penny, W.~D., Trujillo-Barreto, N.~J., and Friston, K.~J. (2005b).
\newblock {Bayesian fMRI time series analysis with spatial priors}.
\newblock {\em NeuroImage}, 24(2):350--362.

\bibitem[Poldrack et~al., 2013]{Poldrack2013}
Poldrack, R.~A., Barch, D.~M., Mitchell, J.~P., Wager, T.~D., Wagner, A.~D.,
  Devlin, J.~T., Cumba, C., Koyejo, O., and Milham, M.~P. (2013).
\newblock {Toward open sharing of task-based fMRI data: the OpenfMRI project.}
\newblock {\em Frontiers in neuroinformatics}, 7:12.

\bibitem[Rad et~al., 2016]{Rad}
Rad, K.~R., Machado, T.~A., and Paninski, L. (2016).
\newblock {Robust and scalable Bayesian analysis of spatial neural tuning
  function data}.
\newblock {\em arXiv:1606.07845v1}.

\bibitem[Risser et~al., 2011]{Risser2011}
Risser, L., Vincent, T., Forbes, F., Idier, J., and Ciuciu, P. (2011).
\newblock {Min-max Extrapolation Scheme for Fast Estimation of 3D Potts Field
  Partition Functions. Application to the Joint Detection-Estimation of Brain
  Activity in fMRI}.
\newblock {\em Journal of Signal Processing Systems}, 65(3):325--338.

\bibitem[Rue and Held, 2005]{Isham2004}
Rue, H. and Held, L. (2005).
\newblock {\em {Gaussian Markov Random Fields: Theory and Applications}}.
\newblock CRC Press.

\bibitem[Rue et~al., 2009]{Solis-Trapala2009}
Rue, H., Martino, S., and Chopin, N. (2009).
\newblock {Approximate Bayesian inference for latent Gaussian models by using
  integrated nested Laplace approximation}.
\newblock {\em Journal of the Royal Statistical Society, Series B},
  71(2):319--392.

\bibitem[Smith and Fahrmeir, 2007]{Smith2007}
Smith, M. and Fahrmeir, L. (2007).
\newblock {Spatial Bayesian variable selection with application to functional
  magnetic resonance imaging}.
\newblock {\em Journal of the American Statistical Association}, 102:417--431.

\bibitem[Sun et~al., 2015]{Sun2015}
Sun, W., Reich, B.~J., {Tony Cai}, T., Guindani, M., and Schwartzman, A.
  (2015).
\newblock {False discovery control in large-scale spatial multiple testing}.
\newblock {\em Journal of the Royal Statistical Society. Series B: Statistical
  Methodology}, 77(1):59--83.

\bibitem[Teng et~al., 2016]{Teng2016}
Teng, M., Johnson, T., and Nathoo, F. (2016).
\newblock {A Comparison of Variational Bayes and Hamiltonian Monte Carlo for
  Bayesian fMRI Time Series Analysis with Spatial Priors}.
\newblock {\em arXiv:1609.02123v1}.

\bibitem[Thirion et~al., 2014]{Thirion2014}
Thirion, B., Varoquaux, G., Dohmatob, E., and Poline, J.~B. (2014).
\newblock {Which fMRI clustering gives good brain parcellations?}
\newblock {\em Frontiers in Neuroscience}, 8:167.

\bibitem[Vincent et~al., 2010]{Vincent2010a}
Vincent, T., Risser, L., and Ciuciu, P. (2010).
\newblock {Spatially Adaptive Mixture Modeling for Analysis of fMRI Time
  Series}.
\newblock {\em IEEE transactions on medical imaging}, 29(4):1059--1074.

\bibitem[Woods, 1972]{Woods1972}
Woods, J.~W. (1972).
\newblock {Two-Dimensional Discrete Markovian Fields}.
\newblock {\em IEEE Transactions on Information Theory}, 18(2):232--240.

\bibitem[Woolrich et~al., 2004]{Woolrich2004c}
Woolrich, M.~W., Jenkinson, M., Brady, J.~M., and Smith, S.~M. (2004).
\newblock {Fully Bayesian spatio-temporal modeling of FMRI data.}
\newblock {\em IEEE transactions on medical imaging}, 23(2):213--31.

\bibitem[Xu et~al., 2009]{Xu2009}
Xu, L., Johnson, T.~D., Nichols, T.~E., and Nee, D.~E. (2009).
\newblock {Modeling inter-subject variability in fMRI activation location: A
  bayesian hierarchical spatial model}.
\newblock {\em Biometrics}, 65(4):1041--1051.

\bibitem[Yue et~al., 2014]{Yue2014}
Yue, Y.~R., Lindquist, M.~a., Bolin, D., Lindgren, F., Simpson, D., and Rue, H.
  (2014).
\newblock {A Bayesian General Linear Modeling Approach to Slice-wise fMRI Data
  Analysis}.
\newblock {\em Preprint}.

\bibitem[Zhang et~al., 2016]{Zhang2016a}
Zhang, L., Guindani, M., Versace, F., Engelmann, J.~M., and Vannucci, M.
  (2016).
\newblock {A spatiotemporal nonparametric Bayesian model of multi-subject fMRI
  data}.
\newblock {\em Ann. Appl. Stat.}, 10(2):638--666.

\bibitem[Zhang et~al., 2014]{Zhang2014}
Zhang, L., Guindani, M., Versace, F., and Vannucci, M. (2014).
\newblock {A spatio-temporal nonparametric Bayesian variable selection model of
  fMRI data for clustering correlated time courses}.
\newblock {\em NeuroImage}, 95:162--175.

\end{thebibliography}

\appendix

\section{Derivation of full conditional posteriors for the MCMC algorithm\label{Appendix full conditionals derivations}}

This section is divided into two pieces, the first handling the case
when the noise in each voxel is modeled as i.i.d. over time and the
second handling the case with auto-regressive temporal noise. The
first part is shorter and basically contains all the concepts needed
for the second part. It should be noted that adding a temporal model
does not add much too the time complexity as long as $P<K$, which
is normally the case. We also provide an expression for computing
the joint posterior ratio.

\subsection{i.i.d. noise model}

\subsubsection*{The likelihood}

The likelihood in (\ref{eq:likelihood}) can be expressed in logs
as
\begin{eqnarray}
\log p\left(\mathbf{Y}|\mathbf{W},\boldsymbol{\lambda}\right) & = & \frac{T}{2}\sum_{n=1}^{N}\log\left(\lambda_{n}\right)\label{eq:like_iid}\\
 &  & -\frac{1}{2}\sum_{n=1}^{N}\lambda_{n}\left[\mathbf{Y}_{\cdot,n}^{\prime}\mathbf{Y}_{\cdot,n}-2\mathbf{Y}_{\cdot,n}^{\prime}\mathbf{X}\mathbf{W}_{\cdot,n}+\mathbf{W}_{\cdot,n}^{\prime}\mathbf{X}^{\prime}\mathbf{X}\mathbf{W}_{\cdot,n}\right]+const,\nonumber 
\end{eqnarray}
where we have omitted everything that is constant with respect to
the parameters. Since $\mathbf{Y}$ and $\mathbf{X}$ are data that
will not change during the MCMC algorithm, quantities as $\mathbf{Y}_{\cdot,n}^{\prime}\mathbf{Y}_{\cdot,n}^{\,}$,
$\mathbf{Y}_{\cdot,n}^{\prime}\mathbf{X}$ and $\mathbf{X}^{\prime}\mathbf{X}$
can be effectively pre-computed, removing the time dimension from
the likelihood which leads to significant speed up. This is similar
to what is done in \citet{Penny2005}.

\subsubsection*{Full conditional posterior of $\mathbf{W}$}

\begin{eqnarray}
\log p\left(\mathbf{W}|\mathbf{Y},\boldsymbol{\lambda},\boldsymbol{\alpha}\right) & = & \log p\left(\mathbf{Y}|\mathbf{W},\boldsymbol{\lambda}\right)+\log p\left(\mathbf{W}|\boldsymbol{\alpha}\right)+const\label{eq:fullcondiidW}\\
 & = & -\frac{1}{2}\left[\sum_{n=1}^{N}\lambda_{n}\left(\mathbf{W}_{\cdot,n}^{\prime}\mathbf{X}^{\prime}\mathbf{X}\mathbf{W}_{\cdot,n}-2\mathbf{Y}_{\cdot,n}^{\prime}\mathbf{X}\mathbf{W}_{\cdot,n}\right)+\sum_{k=1}^{K}\mathbf{W}_{k,\cdot}\alpha_{k}\mathbf{D}_{w}\mathbf{W}_{k,\cdot}^{\prime}\right]+const\nonumber \\
 & = & -\frac{1}{2}\mathbf{w}_{r}^{\prime}\tilde{\mathbf{B}}\mathbf{w}_{r}+\mathbf{b}_{w}^{\prime}\mathbf{w}_{r}+const,\nonumber \\
\mathbf{b}_{w} & = & vec\left(diag\left(\boldsymbol{\lambda}\right)\mathbf{Y}^{\prime}\mathbf{X}\right),\nonumber \\
\tilde{\mathbf{B}} & = & \mathbf{X}^{\prime}\mathbf{X}\otimes diag\left(\boldsymbol{\lambda}\right)+diag\left(\boldsymbol{\alpha}\right)\otimes\mathbf{D}_{w},\nonumber 
\end{eqnarray}
or equivalently $\mathbf{w}_{r}|\mathbf{Y},\boldsymbol{\lambda},\boldsymbol{\alpha}\sim\mathcal{N}\left(\tilde{\mathbf{B}}^{-1}\mathbf{b}_{w},\tilde{\mathbf{B}}^{-1}\right)$,
where $\mathbf{w}_{r}=vec\left(\mathbf{W}'\right)$.

\subsubsection*{Full conditional posterior of $\boldsymbol{\lambda}$}

\begin{eqnarray}
\log p\left(\boldsymbol{\lambda}|\mathbf{Y},\mathbf{W},\boldsymbol{\alpha}\right) & = & \log p\left(\mathbf{Y}|\mathbf{W},\boldsymbol{\lambda}\right)+\log p\left(\boldsymbol{\lambda}\right)+const\label{eq:fullcondiidlambda}\\
 & = & \frac{T}{2}\sum_{n=1}^{N}\log\left(\lambda_{n}\right)-\frac{1}{2}\sum_{n=1}^{N}\lambda_{n}\left[\mathbf{Y}_{\cdot,n}^{\prime}\mathbf{Y}_{\cdot,n}-2\mathbf{Y}_{\cdot,n}^{\prime}\mathbf{X}\mathbf{W}_{\cdot,n}^{\prime}+\mathbf{W}_{\cdot,n}^{\prime}\mathbf{X}^{\prime}\mathbf{X}\mathbf{W}_{\cdot,n}\right]\nonumber \\
 &  & +\left(u_{2}-1\right)\sum_{n=1}^{N}\log\left(\lambda_{n}\right)-\sum_{n=1}^{N}\frac{\lambda_{n}}{u_{1}}+const\nonumber \\
 & = & \left(\tilde{u}_{2}-1\right)\sum_{n=1}^{N}\log\left(\lambda_{n}\right)-\sum_{n=1}^{N}\frac{\lambda_{n}}{\tilde{u}_{1n}}+const,\nonumber \\
\frac{1}{\tilde{u}_{1n}} & = & \frac{1}{2}\left(\mathbf{Y}_{\cdot,n}^{\prime}\mathbf{Y}_{\cdot,n}-2\mathbf{Y}_{\cdot,n}^{\prime}\mathbf{X}\mathbf{W}_{\cdot,n}^{\prime}+\mathbf{W}_{\cdot,n}^{\prime}\mathbf{X}^{\prime}\mathbf{X}\mathbf{W}_{\cdot,n}\right)+\frac{1}{u_{1}},\nonumber \\
\tilde{u}_{2} & = & \frac{T}{2}+u_{2},\nonumber 
\end{eqnarray}
so $\lambda_{n}|\mathbf{Y},\mathbf{W}\sim Ga\left(\tilde{u}_{1n},\tilde{u}_{2}\right)$
for all $n$.

\subsubsection*{Full conditional posterior of $\boldsymbol{\alpha}$}

\begin{eqnarray}
\log p\left(\boldsymbol{\alpha}|\mathbf{Y},\mathbf{W},\boldsymbol{\lambda}\right) & = & \log p\left(\mathbf{Y}|\mathbf{W},\boldsymbol{\lambda}\right)+\log p\left(\mathbf{W}|\boldsymbol{\alpha}\right)+\log p\left(\boldsymbol{\alpha}\right)+const\label{eq:fullcondiidalpha}\\
 & = & \frac{N}{2}\sum_{k=1}^{K}\log\left(\alpha_{k}\right)-\frac{1}{2}\sum_{k=1}^{K}\mathbf{W}_{k,\cdot}\alpha_{k}\mathbf{D}_{w}\mathbf{W}_{k,\cdot}^{\prime}\nonumber \\
 &  & +\left(q_{2}-1\right)\sum_{k=1}^{K}\log\left(\alpha_{k}\right)-\sum_{k=1}^{K}\frac{\alpha_{k}}{q_{1}}+const\nonumber \\
 & = & \left(\tilde{q}_{2}-1\right)\sum_{k=1}^{K}\log\left(\alpha_{k}\right)-\sum_{k=1}^{K}\frac{\alpha_{k}}{\tilde{q}_{1k}}+const,\nonumber \\
\frac{1}{\tilde{q}_{1k}} & = & \frac{1}{2}\mathbf{W}_{k,\cdot}\mathbf{D}_{w}\mathbf{W}_{k,\cdot}^{\prime}+\frac{1}{q_{1}},\nonumber \\
\tilde{q}_{2} & = & \frac{N}{2}+q_{2},\nonumber 
\end{eqnarray}
so $\alpha_{k}|\mathbf{W}\sim Ga\left(\tilde{q}_{1k},\tilde{q}_{2}\right)$
for all $k$.

\subsection{Temporal noise model}

~\\
The derivations of the full conditionals for the temporal noise model
follows the same pattern as for the i.i.d. model, why we leave out
some steps for brevity. Also note that the form of the posterior for
the AR coefficients $\mathbf{A}$ will be very similar to that of
the regression coefficients $\mathbf{W}$, and the hyperparameter
$\boldsymbol{\beta}$ has the same form as $\boldsymbol{\alpha}$.
The permutation matrices $\mathbf{H}_{w}$ and $\mathbf{H}_{a}$ are
defined as in \citet{Penny2007} such that $vec\left(\mathbf{W}\right)=\mathbf{H}_{w}vec\left(\mathbf{W}^{\prime}\right)$
and $vec\left(\mathbf{A}\right)=\mathbf{H}_{a}vec\left(\mathbf{A}^{\prime}\right)$. 

\subsubsection*{The likelihood}

Using the temporal noise model, the likelihood can be expressed as
\begin{equation}
p\left(\mathbf{Y}|\mathbf{W},\mathbf{A},\boldsymbol{\lambda}\right)\propto\prod_{t=P+1}^{T}\prod_{n=1}^{N}\mathcal{N}\left(Y_{tn}-\mathbf{X}_{t,\cdot}\mathbf{W}_{\cdot,n};\left(\mathbf{d}_{tn}-\tilde{\mathbf{X}}_{t}\mathbf{W}_{\cdot,n}\right)^{\prime}\mathbf{A}_{\cdot,n},\lambda_{n}^{-1}\right),\label{eq:likeliTemp}
\end{equation}
where $\tilde{\mathbf{X}}_{t}$ is a $P\times K$ matrix containing
the $P$ rows of the design matrix prior to time point $t$ and $\mathbf{d}_{tn}$
is $P\times1$ and similarly contains the $P$ values of $\mathbf{Y}_{\cdot,n}$
just before $t$. Note that we condition on the first $P$ time points
for simplicity. The non-constant part of the log-likelihood can be
written as
\begin{eqnarray}
\log p\left(\mathbf{Y}|\cdot\right) & = & \frac{T-P}{2}\sum_{n=1}^{N}\log\left(\lambda_{n}\right)-\frac{1}{2}\sum_{t=P+1}^{T}\sum_{n=1}^{N}\left[\left(Y_{tn}-\mathbf{X}_{t,\cdot}\mathbf{W}_{\cdot,n}\right)-\left(\mathbf{d}_{tn}-\tilde{\mathbf{X}}_{t}\mathbf{W}_{\cdot,n}\right)^{\prime}\mathbf{A}_{\cdot,n}\right]^{\prime}\lambda_{n}\nonumber \\
 &  & \times\left[\left(Y_{tn}-\mathbf{X}_{t,\cdot}\mathbf{W}_{\cdot,n}\right)-\left(\mathbf{d}_{tn}-\tilde{\mathbf{X}}_{t}\mathbf{W}_{\cdot,n}\right)^{\prime}\mathbf{A}_{\cdot,n}\right]+const\nonumber \\
 & = & \frac{T-P}{2}\sum_{n=1}^{N}\log\left(\lambda_{n}\right)-\frac{1}{2}\sum_{n=1}^{N}\left[\left(\mathbf{Y}_{\cdot,n}-\mathbf{X}\mathbf{W}_{\cdot,n}\right)-\left(\mathbf{d}_{n}-\tilde{\mathbf{X}}\mathbf{W}_{\cdot,n}\right)^{\prime}\mathbf{A}_{\cdot,n}\right]^{\prime}\lambda_{n}\nonumber \\
 &  & \times\left[\left(\mathbf{Y}_{\cdot,n}-\mathbf{X}\mathbf{W}_{\cdot,n}\right)-\left(\mathbf{d}_{n}-\tilde{\mathbf{X}}\mathbf{W}_{\cdot,n}\right)^{\prime}\mathbf{A}_{\cdot,n}\right]+const.\label{eq:likeliTemp2}
\end{eqnarray}
In the last expression the sums and indexing with respect to $t$
has been removed. Since none of the parameters depend on the time
dimension, this expression can be rewritten so that sums and matrix
multiplications over the time dimension can be isolated to the data
$\left(\mathbf{Y},\mathbf{X},\mathbf{d},\tilde{\mathbf{X}}\right)$,
so that they can be pre-computed outside the Gibbs algorithm, leading
to a higher computational efficiency as in \citet{Penny2005}. Note
that the size of $\mathbf{d}_{n}$ is $P\times\left(T-P\right)$ and
the size of $\tilde{\mathbf{X}}$ is $P\times\left(T-P\right)\times K$.
Matrix multiplications including the 3-dimensional matrix $\tilde{\mathbf{X}}$
and other tensors will be carried out over the appropriate dimension
in what follows, even if this is not stated explicitly. The log-likelihood
can now be rewritten as
\begin{eqnarray}
\log p\left(\mathbf{Y}|\cdot\right) & = & \frac{T-P}{2}\sum_{n=1}^{N}\log\left(\lambda_{n}\right)\label{eq:log-like}\\
 &  & -\frac{1}{2}\sum_{n=1}^{N}\lambda_{n}\left[\mathbf{Y}_{\cdot,n}^{\prime}\mathbf{Y}_{\cdot,n}-2\mathbf{Y}_{\cdot,n}^{\prime}\mathbf{X}\mathbf{W}_{\cdot,n}+\mathbf{W}_{\cdot,n}^{\prime}\mathbf{X}^{\prime}\mathbf{X}\mathbf{W}_{\cdot,n}-2\mathbf{Y}_{\cdot,n}^{\prime}\mathbf{d}_{n}^{\prime}\mathbf{A}_{\cdot,n}\right.\nonumber \\
 &  & +\mathbf{A}_{\cdot,n}^{\prime}\mathbf{d}_{n}\mathbf{d}_{n}^{\prime}\mathbf{A}_{\cdot,n}+\mathbf{W}_{\cdot,n}^{\prime}\mathbf{B}_{n}^{\prime}\mathbf{A}_{\cdot,n}+\mathbf{A}_{\cdot,n}^{\prime}\mathbf{B}_{n}\mathbf{W}_{\cdot,n}-\mathbf{W}_{\cdot,n}^{\prime}\left(\mathbf{R}\mathbf{A}_{\cdot,n}+\left(\mathbf{R}\mathbf{A}_{\cdot,n}\right)^{\prime}\right)\mathbf{W}_{\cdot,n}\nonumber \\
 &  & \left.-\mathbf{A}_{\cdot,n}^{\prime}\left(\mathbf{D}_{n}\mathbf{W}_{\cdot,n}+\left(\mathbf{D}_{n}\mathbf{W}_{\cdot,n}\right)^{\prime}\right)\mathbf{A}_{\cdot,n}+\mathbf{W}_{\cdot,n}^{\prime}\left(\mathbf{A}_{\cdot,n}^{\prime}\mathbf{S}\mathbf{A}_{\cdot,n}\right)\mathbf{W}_{\cdot,n}\right]+const,\nonumber 
\end{eqnarray}
where 
\begin{eqnarray*}
\mathbf{B}_{n} & =\mathbf{Y}_{\cdot,n}^{\prime}\tilde{\mathbf{X}}+\mathbf{d}_{n}\mathbf{X} & \text{is of size }P\times K,\\
\mathbf{R} & =\mathbf{X}^{\prime}\tilde{\mathbf{X}} & \text{is of size }K\times K\times P,\\
\mathbf{D}_{n} & =\mathbf{d}_{n}\tilde{\mathbf{X}} & \text{is of size }P\times K\times P,\\
\mathbf{S} & =\tilde{\mathbf{X}}\tilde{\mathbf{X}} & \text{is of size }P\times K\times K\times P.
\end{eqnarray*}

\subsubsection*{Full conditional posterior of $\mathbf{W}$}

\begin{eqnarray}
\log p\left(\mathbf{W}|\mathbf{Y},\cdot\right) & = & -\frac{1}{2}\mathbf{w}_{r}^{\prime}\tilde{\mathbf{B}}\mathbf{w}_{r}+\mathbf{b}_{w}^{\prime}\mathbf{w}_{r}+const,\label{eq:full cond W temporal}\\
\mathbf{b}_{w} & = & vec\left(\left[\begin{array}{c}
\vdots\\
\lambda_{n}\left(\mathbf{Y}_{\cdot,n}^{\prime}\mathbf{X}-\mathbf{A}_{\cdot,n}^{\prime}\mathbf{B}_{n}+\mathbf{A}_{\cdot,n}^{\prime}\mathbf{D}_{n}\mathbf{A}_{\cdot,n}\right)\\
\vdots
\end{array}\right]_{n\in\left\{ 1,\ldots,N\right\} }\right),\nonumber \\
\tilde{\mathbf{B}} & = & \mathbf{H}_{w}^{\prime}\underset{n\in\left\{ 1,\ldots,N\right\} }{blkdiag}\left[\lambda_{n}\left(\mathbf{X}^{\prime}\mathbf{X}-\mathbf{R}\mathbf{A}_{\cdot,n}-\left(\mathbf{R}\mathbf{A}_{\cdot,n}\right)^{\prime}+\mathbf{A}_{\cdot,n}^{\prime}\mathbf{S}\mathbf{A}_{\cdot,n}\right)\right]\mathbf{H}_{w}\nonumber \\
 &  & +diag\left(\boldsymbol{\alpha}\right)\otimes\mathbf{D}_{w},\nonumber 
\end{eqnarray}
where $\underset{n\in\left\{ 1,\ldots,N\right\} }{blkdiag}\left[\mathbf{C}_{n}\right]$
is a $KN\times KN$ block diagonal matrix with the $K\times K$ matrix
$\mathbf{C}_{n}$ as the $n$th block. So $\mathbf{w}_{r}|\mathbf{Y},\cdot\sim\mathcal{N}\left(\tilde{\mathbf{B}}^{-1}\mathbf{b}_{w},\tilde{\mathbf{B}}^{-1}\right).$

\subsubsection*{Full conditional posterior of $\mathbf{A}$}

\begin{eqnarray}
\log p\left(\mathbf{A}|\mathbf{Y},\cdot\right) & = & -\frac{1}{2}\mathbf{a}_{r}^{\prime}\tilde{\mathbf{J}}\mathbf{a}_{r}+\mathbf{b}_{a}^{\prime}\mathbf{a}_{r}+const,\label{eq:fullcondTempA}\\
\mathbf{b}_{a} & = & vec\left(\left[\begin{array}{c}
\vdots\\
\lambda_{n}\left(\mathbf{Y}_{\cdot,n}^{\prime}\mathbf{d}_{n}^{\prime}-\mathbf{W}_{\cdot,n}^{\prime}\mathbf{B}_{n}^{\prime}+\mathbf{W}_{\cdot,n}^{\prime}\mathbf{R}\mathbf{W}_{\cdot,n}\right)\\
\vdots
\end{array}\right]_{n\in\left\{ 1,\ldots,N\right\} }\right),\nonumber \\
\tilde{\mathbf{J}} & = & \mathbf{H}_{a}^{\prime}\underset{n\in\left\{ 1,\ldots,N\right\} }{blkdiag}\left[\lambda_{n}\left(\mathbf{d}_{n}\mathbf{d}_{n}^{\prime}-\mathbf{D}_{n}\mathbf{W}_{\cdot,n}-\left(\mathbf{D}_{n}\mathbf{W}_{\cdot,n}\right)^{\prime}+\mathbf{W}_{\cdot,n}^{\prime}\mathbf{S}\mathbf{W}_{\cdot,n}\right)\right]\mathbf{H}_{a}\nonumber \\
 &  & +diag\left(\boldsymbol{\beta}\right)\otimes\mathbf{D}_{a},\nonumber 
\end{eqnarray}
so $\mathbf{a}_{r}|\mathbf{Y},\cdot\sim\mathcal{N}\left(\tilde{\mathbf{J}}^{-1}\mathbf{b}_{a},\tilde{\mathbf{J}}^{-1}\right).$

\subsubsection*{Full conditional posterior of $\boldsymbol{\lambda}$}

\begin{eqnarray}
\log p\left(\boldsymbol{\lambda}|\mathbf{Y},\cdot\right) & = & \left(\tilde{u}_{2}-1\right)\sum_{n=1}^{N}\log\left(\lambda_{n}\right)-\sum_{n=1}^{N}\frac{\lambda_{n}}{\tilde{u}_{1n}}+const,\label{eq:fullcondTemplambda}\\
\frac{1}{\tilde{u}_{1n}} & = & \frac{1}{u_{1}}+\frac{1}{2}\left[\mathbf{Y}_{\cdot,n}^{\prime}\mathbf{Y}_{\cdot,n}-2\mathbf{Y}_{\cdot,n}^{\prime}\mathbf{X}\mathbf{W}_{\cdot,n}+\mathbf{W}_{\cdot,n}^{\prime}\mathbf{X}^{\prime}\mathbf{X}\mathbf{W}_{\cdot,n}-2\mathbf{Y}_{\cdot,n}^{\prime}\mathbf{d}_{n}^{\prime}\mathbf{A}_{\cdot,n}\right.\nonumber \\
 &  & +\mathbf{A}_{\cdot,n}^{\prime}\mathbf{d}_{n}\mathbf{d}_{n}^{\prime}\mathbf{A}_{\cdot,n}+\mathbf{W}_{\cdot,n}^{\prime}\mathbf{B}_{n}^{\prime}\mathbf{A}_{\cdot,n}+\mathbf{A}_{\cdot,n}^{\prime}\mathbf{B}_{n}\mathbf{W}_{\cdot,n}-\mathbf{W}_{\cdot,n}^{\prime}\left(\mathbf{R}\mathbf{A}_{\cdot,n}+\left(\mathbf{R}\mathbf{A}_{\cdot,n}\right)^{\prime}\right)\mathbf{W}_{\cdot,n}\nonumber \\
 &  & \left.-\mathbf{A}_{\cdot,n}^{\prime}\left(\mathbf{D}_{n}\mathbf{W}_{\cdot,n}+\left(\mathbf{D}_{n}\mathbf{W}_{\cdot,n}\right)^{\prime}\right)\mathbf{A}_{\cdot,n}+\mathbf{W}_{\cdot,n}^{\prime}\left(\mathbf{A}_{\cdot,n}^{\prime}\mathbf{S}\mathbf{A}_{\cdot,n}\right)\mathbf{W}_{\cdot,n}\right],\nonumber \\
\tilde{u}_{2} & = & \frac{T-P}{2}+u_{2},\nonumber 
\end{eqnarray}
so $\lambda_{n}|\mathbf{Y},\cdot\sim Ga\left(\tilde{u}_{1n},\tilde{u}_{2}\right)$
for all $n$.

\subsubsection*{Full conditional posterior of $\boldsymbol{\alpha}$}

\begin{eqnarray}
\log p\left(\boldsymbol{\alpha}|\cdot\right) & = & \left(\tilde{q}_{2}-1\right)\sum_{k=1}^{K}\log\left(\alpha_{k}\right)-\sum_{k=1}^{K}\frac{\alpha_{k}}{\tilde{q}_{1k}}+const,\label{eq:fullcondTempa}\\
\frac{1}{\tilde{q}_{1k}} & = & \frac{1}{2}\mathbf{W}_{k,\cdot}\mathbf{D}_{w}\mathbf{W}_{k,\cdot}^{\prime}+\frac{1}{q_{1}},\nonumber \\
\tilde{q}_{2} & = & \frac{N}{2}+q_{2},\nonumber 
\end{eqnarray}
so $\alpha_{k}|\cdot\sim Ga\left(\tilde{q}_{1k},\tilde{q}_{2}\right)$
for all $k$. This is exactly the same as for the i.i.d. case.

\subsubsection*{Full conditional posterior of $\boldsymbol{\beta}$}

\begin{eqnarray}
\log p\left(\boldsymbol{\beta}|\cdot\right) & = & \left(\tilde{r}_{2}-1\right)\sum_{p=1}^{P}\log\left(\beta_{p}\right)-\sum_{p=1}^{P}\frac{\beta_{p}}{\tilde{r}_{1p}}+const,\label{eq:fullcondTempbeta}\\
\frac{1}{\tilde{r}_{1p}} & = & \frac{1}{2}\mathbf{A}_{p,\cdot}\mathbf{D}_{a}\mathbf{A}_{p,\cdot}^{\prime}+\frac{1}{r_{1}},\nonumber \\
\tilde{r}_{2} & = & \frac{N}{2}+r_{2},\nonumber 
\end{eqnarray}
so $\beta_{p}|\cdot\sim Ga\left(\tilde{r}_{1p},\tilde{r}_{2}\right)$
for all $p$.

\subsubsection*{Joint posterior ratio}

The ratio of the joint posterior $p\left(\mathbf{W},\mathbf{A},\boldsymbol{\lambda},\boldsymbol{\alpha},\boldsymbol{\beta}|\mathbf{Y}\right)$
evaluated for two different sets of parameter values, $\left\{ \mathbf{W},\mathbf{A},\boldsymbol{\lambda},\boldsymbol{\alpha},\boldsymbol{\beta}\right\} $
and $\left\{ \mathbf{W}^{*},\mathbf{A}^{*},\boldsymbol{\lambda}^{*},\boldsymbol{\alpha}^{*},\boldsymbol{\beta}^{*}\right\} $,
can be used to compare the posterior density in different points,
even when the normalized joint posterior itself is not available in
closed form. The ratio can be computed as the ratio of the unnormalized
joint posterior $\tilde{p}\left(\mathbf{W},\mathbf{A},\boldsymbol{\lambda},\boldsymbol{\alpha},\boldsymbol{\beta}|\mathbf{Y}\right)$
defined by
\begin{eqnarray}
\log\tilde{p}\left(\mathbf{W},\mathbf{A},\boldsymbol{\lambda},\boldsymbol{\alpha},\boldsymbol{\beta}|\mathbf{Y}\right) & = & \log\tilde{p}\left(\mathbf{Y}|\cdot\right)+\log\tilde{p}\left(\mathbf{W}|\boldsymbol{\alpha}\right)+\log\tilde{p}\left(\mathbf{A}|\boldsymbol{\beta}\right)+\log p\left(\boldsymbol{\lambda}\right)\label{eq:UnnormalisedJointPosterior}\\
 &  & +\log p\left(\boldsymbol{\alpha}\right)+\log p\left(\boldsymbol{\beta}\right),\nonumber 
\end{eqnarray}
where $\log\tilde{p}\left(\mathbf{Y}|\cdot\right)$ is defined as
in equation (\ref{eq:log-like}) without the constant part, 
\begin{eqnarray}
\log\tilde{p}\left(\mathbf{W}|\boldsymbol{\alpha}\right) & = & \sum_{k=1}^{K}\left[\log\left(\left|\alpha_{k}\mathbf{D}_{w}\right|^{1/2}\right)-\frac{1}{2}\alpha_{k}\mathbf{W}_{k,\cdot}\mathbf{D}_{w}\mathbf{W}_{k,\cdot}^{\prime}\right]+const\label{eq:UnnormalisedWPrior}\\
 & = & \sum_{k=1}^{K}\left[\frac{N}{2}\log\left(\alpha_{k}\right)-\frac{1}{2}\alpha_{k}\mathbf{W}_{k,\cdot}\mathbf{D}_{w}\mathbf{W}_{k,\cdot}^{\prime}\right]\nonumber 
\end{eqnarray}
and $\log\tilde{p}\left(\mathbf{A}|\boldsymbol{\beta}\right)$ is
defined correspondingly.

\section{Derivation of Spatial VB posteriors\label{Appendix NFVB derivations}}

From equation (\ref{eq:Bishops vb posterior}) one sees that the SVB
approximate posteriors can be derived from the full conditionals in
Appendix \ref{Appendix full conditionals derivations}, since
\begin{equation}
\log q\left(\theta_{j}\right)=E_{\theta_{-j}}\left[\log p\left(\mathbf{Y},\boldsymbol{\theta}\right)\right]+const=E_{\theta_{-j}}\left[\log p\left(\theta_{j}|\mathbf{Y},\theta_{-j}\right)\right]+const.\label{eq:NFVBupdate}
\end{equation}
The only difference is that we must take the expectation with respect
to all other parameters. It turns out that the dependencies on $\boldsymbol{\lambda},\boldsymbol{\alpha}$
and $\boldsymbol{\beta}$ are always linear, so we can just replace
these with their expectations $\bar{\boldsymbol{\lambda}},\bar{\boldsymbol{\alpha}}$
and $\bar{\boldsymbol{\beta}}$ under respective SVB gamma posterior.
For example, 
\begin{equation}
\bar{\alpha}_{k}=E_{\boldsymbol{\alpha}}\left[\alpha_{k}\right]=\tilde{q}_{1k}^{SVB}\cdot\tilde{q}_{2}^{SVB}.\label{eq:alphamean}
\end{equation}
For dependencies on $\mathbf{W}$ and $\mathbf{A}$ we use MC approximations
as in (\ref{eq:MC approx}) to compute the expectations, by simulating
$N_{s}$ samples of $\mathbf{W}$ and $\mathbf{A}$ respectively.
For brevity, we do not derive all approximate posteriors here, only
$q\left(\mathbf{W}\right)$ for the temporal model as an example.
\begin{eqnarray}
\log q\left(\mathbf{w}_{r}\right) & = & E_{\mathbf{A},\boldsymbol{\lambda},\boldsymbol{\alpha},\boldsymbol{\beta}}\left[\log p\left(\mathbf{W}|\mathbf{Y},\cdot\right)\right]+const\label{eq:NFVBw}\\
 & = & -\frac{1}{2}\mathbf{w}_{r}^{\prime}E_{\mathbf{A},\boldsymbol{\lambda},\boldsymbol{\alpha},\boldsymbol{\beta}}\left[\tilde{\mathbf{B}}\right]\mathbf{w}_{r}+E_{\mathbf{A},\boldsymbol{\lambda},\boldsymbol{\alpha},\boldsymbol{\beta}}\left[\mathbf{b}_{w}\right]\mathbf{w}_{r}+const,\nonumber 
\end{eqnarray}
with $\tilde{\mathbf{B}}$ and $\mathbf{b}_{w}$ from equation (\ref{eq:full cond W temporal}).
The expectations are computed as 
\begin{eqnarray}
\mathbf{b}_{w}^{SVB} & = & E_{\mathbf{A},\boldsymbol{\lambda},\boldsymbol{\alpha},\boldsymbol{\beta}}\left[\mathbf{b}_{w}\right]\label{eq:NFVBbB}\\
 & = & \frac{1}{N_{s}}\sum_{j=1}^{N_{s}}vec\left(\left[\begin{array}{c}
\vdots\\
\bar{\lambda}_{n}\left(\mathbf{Y}_{\cdot,n}^{\prime}\mathbf{X}-\mathbf{A}_{\cdot,n}^{\left(j\right)\prime}\mathbf{B}_{n}+\mathbf{A}_{\cdot,n}^{\left(j\right)\prime}\mathbf{D}_{n}\mathbf{A}_{\cdot,n}^{\left(j\right)}\right)\\
\vdots
\end{array}\right]_{n\in\left\{ 1,\ldots,N\right\} }\right)\nonumber \\
\tilde{\mathbf{B}}^{SVB} & = & E_{\mathbf{A},\boldsymbol{\lambda},\boldsymbol{\alpha},\boldsymbol{\beta}}\left[\tilde{\mathbf{B}}\right]\nonumber \\
 & = & \frac{1}{N_{s}}\sum_{j=1}^{N_{s}}\mathbf{H}_{w}^{\prime}\underset{n\in\left\{ 1,\ldots,N\right\} }{blkdiag}\left[\bar{\lambda}_{n}\left(\mathbf{X}^{\prime}\mathbf{X}-\mathbf{R}\mathbf{A}_{\cdot,n}^{\left(j\right)}-\left(\mathbf{R}\mathbf{A}_{\cdot,n}^{\left(j\right)}\right)^{\prime}+\mathbf{A}_{\cdot,n}^{\left(j\right)\prime}\mathbf{S}\mathbf{A}_{\cdot,n}^{\left(j\right)}\right)\right]\mathbf{H}_{w}\nonumber \\
 &  & +diag\left(\bar{\boldsymbol{\alpha}}\right)\otimes\mathbf{D}_{w}.
\end{eqnarray}
So $q\left(\mathbf{W}\right)\sim\mathcal{N}\left(\left(\tilde{\mathbf{B}}^{SVB}\right)^{-1}\mathbf{b}_{w}^{SVB},\left(\tilde{\mathbf{B}}^{SVB}\right)^{-1}\right).$

\section{convergence and implementation details\label{appendix:convergence}}

\begin{figure}
\subfloat[\label{fig:hyperConvergence a}]{\includegraphics[width=0.5\linewidth]{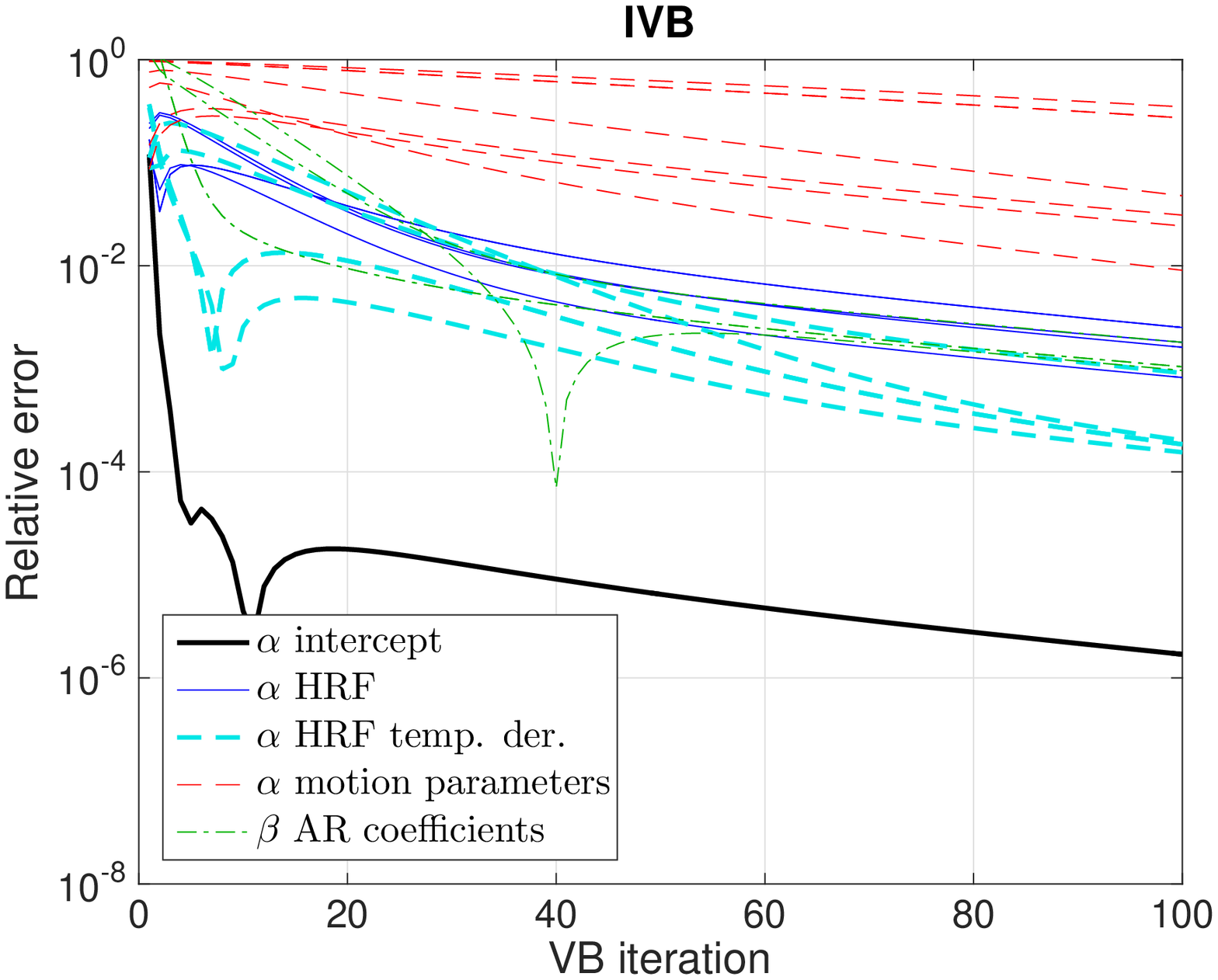}}\subfloat[\label{fig:hyperConvergence b}]{\includegraphics[width=0.5\linewidth]{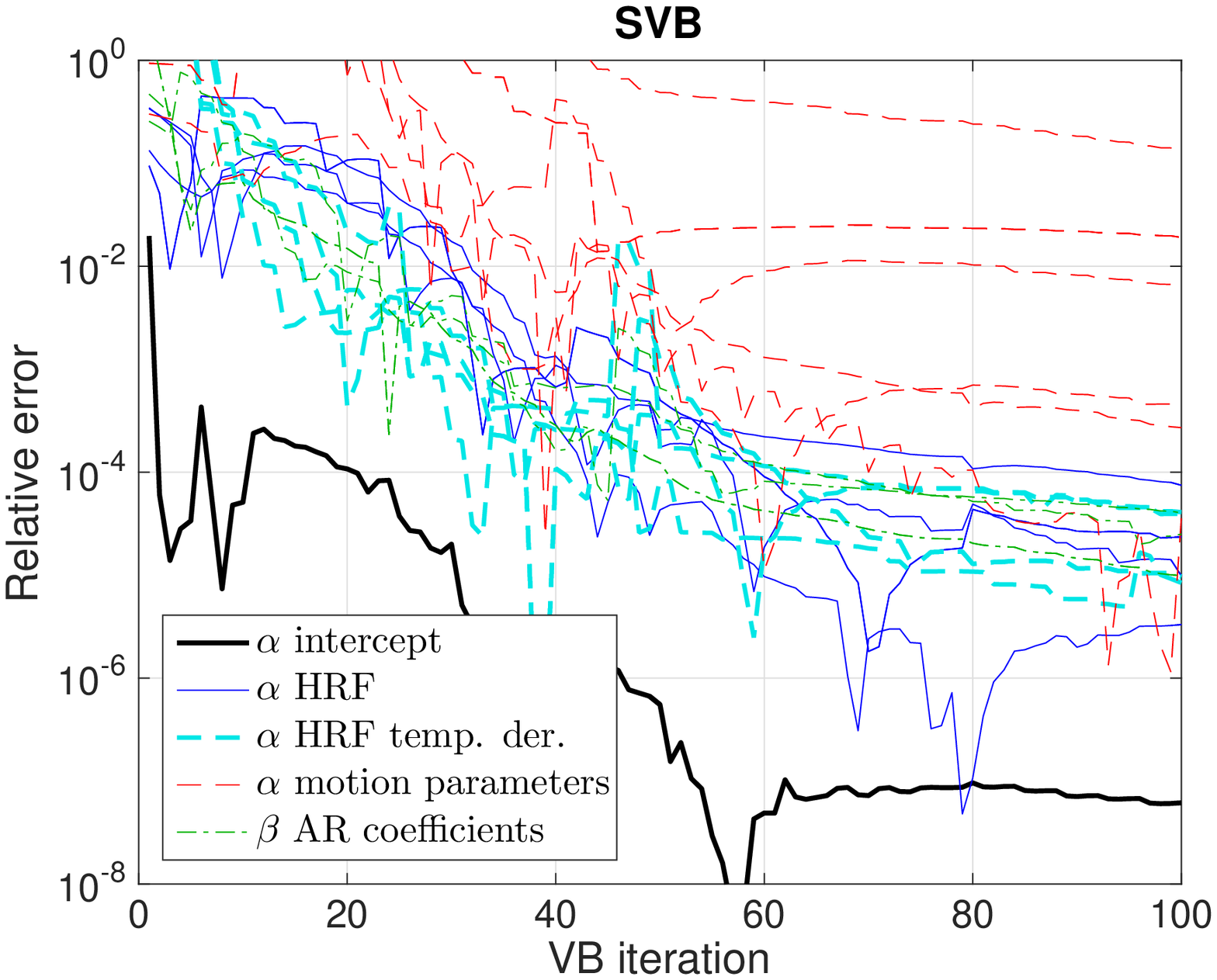}}

\caption{Spatial hyperparameter relative errors by VB iteration number, relative
to the value after 200 iterations, for the face repetition data. The
errors for IVB are shown in (a), while the errors for SVB are shown
in (b). The regression coefficient hyperparameters $\boldsymbol{\alpha}$
are divided by regressor type and $\boldsymbol{\beta}$ are the hyperparameters
for the AR coefficients.\label{fig:hyperConvergence}}
\end{figure}

Our experience from running the MCMC and VB methods is that the convergence
of the algorithms is largely governed by that of the spatial hyperparameters
$\boldsymbol{\alpha}$ and $\boldsymbol{\beta}$. Figure \ref{fig:hyperConvergence a}
shows the relative error by iteration number for the hyperparameters
for IVB, for the presented slice of the face repetition data. The
relative error is here defined as compared to the final value after
a long run (200 iterations) and if $\bar{\alpha}_{k}^{\left[j\right]}$
denotes the value of $\bar{\alpha}_{k}$ after $j$ iterations, then
the relative error $\epsilon_{k}=\left|\nicefrac{\bar{\alpha}_{k}^{\left[j\right]}}{\bar{\alpha}_{k}^{\left[200\right]}}-1\right|$
for the $k$th parameter. We see that $\bar{\alpha}_{k}$ converges
the fastest for the intercept, a bit slower for regressors connected
to the HRF and its temporal derivative and for the AR coefficients,
and the slowest for the head motion nuisance regressors. 

To understand why the parameters differ in convergence speed one has
to consider how the VB algorithm works, updating the approximate posterior
for $\mathbf{W}$ given $\boldsymbol{\alpha}$ and vice versa. Since
$\boldsymbol{\alpha}$ controls the smoothness and shrinkage of $\mathbf{W}$,
there will be much dependence between the estimated posteriors of
these two parameters, leading to slower convergence. In general, the
more informative the data are the faster is seemingly the convergence.
For example, for $k$ corresponding to the intercept in every voxel,
$\mathbf{W}_{k,\cdot}$ is quite well defined from the data, so it
will depend relatively little on $\alpha_{k}$. Given the smoothness
of $\mathbf{W}_{k,\cdot}$, $\alpha_{k}$ is also well determined,
so convergence will be quick. On the other hand, for $k$ corresponding
to head motion, $\mathbf{W}_{k,\cdot}$ will be non-significant in
most voxels, giving more dependence with $\boldsymbol{\alpha}$ and
slower convergence.

It takes almost 50 iterations for all the HRF regressor $\alpha_{k}$:s
(which are the most interesting ones for PPMs) to reach a relative
error smaller than $1\%$ for the IVB algorithm. This is interesting,
since the SPM12 default setting is to use 4 VB iterations and the
other available stopping criterion, based on the model evidence lower
bound, results in 8 iterations. Even though this might seem as too
few, the effect of this on the PPMs is not necessarily so large (see
Figure \ref{fig:faceRepPPMs} and \ref{fig:openfMRIPPMs}), so keeping
the number of iterations low might well be a reasonable strategy in
order to reduce the processing time.

The SVB convergence plot would look similar to that of IVB, if implemented
directly as it is presented above. However, we added some ad hoc steps
to the SVB algorithm to speed up convergence. In contrast to IVB,
the posterior update step for $q\left(\mathbf{W}\right)$ only depends
on the other parameters (in particular on $\boldsymbol{\alpha}$)
and the data, but not on the previous iteration value of $q\left(\mathbf{W}\right)$.
Therefore, if there is some faster way to approach the optimal value
of $\bar{\boldsymbol{\alpha}}$, than the SVB update equation for
$q\left(\boldsymbol{\alpha}\right)$, that speedup will transfer to
$q\left(\mathbf{W}\right)$ as well. In the SVB implementation we
use three ad hoc tricks to speed up convergence, which we have noticed
work well in practice:
\begin{enumerate}
\item In every other VB iteration, instead of accepting the $\tilde{q}_{1k}^{SVB\left[j\right]}$
(and hence $\bar{\alpha}_{k}^{\left[j\right]}=\tilde{q}_{1k}^{SVB\left[j\right]}\cdot\tilde{q}_{2}^{SVB}$)
from the VB update equation, we fit a quadratic function to the values
of the last three iterations, $\bar{\alpha}_{k}^{\left[j-2\right]}$,
$\bar{\alpha}_{k}^{\left[j-1\right]}$ and $\bar{\alpha}_{k}^{\left[j\right]}$,
as a function of the iteration number $j$. If the vertex of the quadratic
function occurs after $j$, we take the value of the function at the
vertex as a prediction of what $\bar{\alpha}_{k}$ will converge to.
Otherwise, we assume that we are far from the final value and set
the prediction to $\bar{\alpha}_{k}^{\left[j-1\right]}+20\left(\bar{\alpha}_{k}^{\left[j\right]}-\bar{\alpha}_{k}^{\left[j-1\right]}\right).$
We set $\bar{\alpha}_{k}^{\left[j\right]}$ to the predicted value
but only allow changes in alpha up to a factor $5$ from the value
proposed by the VB update equation. This methodology will make the
SVB algorithm behave less stable, but allow for larger steps when
$\bar{\alpha}_{k}$ is far from the final value which leads to faster
convergence, especially when data are not so informative.
\item We use the prior mean as starting values for all parameters, which
is a little different from what SPM's IVB does, but leads to faster
convergence for the SVB algorithm.
\item In the first $10$ iterations of the SVB algorithm we use $N_{s}=5$,
which will make these iterations quicker, but not as exact.
\end{enumerate}
The effect of using these tricks can be seen in Figure \ref{fig:hyperConvergence b},
showing faster convergence for the hyperparameters using SVB as compared
to IVB.

Different random seeds in the MC approximation (equation (\ref{eq:MC approx}))
make the SVB algorithm converge to slightly different posteriors.
We reran SVB several times with $N_{s}=100$ and different seeds,
and found very small differences in the results for the simulated
data, but for some real data sets we found that the differences could
be slightly larger. SVB appears to sometimes get stuck in local modes
of the model evidence lower bound that is implicitly optimized by
the VB algorithm. The resulting SVB posteriors for all of these modes
were however always closer to the exact MCMC posterior than the IVB
posterior was. Since the model evidence lower bound is computationally
intractable, we instead compared the different modes using the joint
posterior ratio in equation (\ref{eq:UnnormalisedJointPosterior})
evaluated in the SVB posterior mean to decide which results to present
for these problematic data sets. The same multimodal issues were not
observed when rerunning MCMC with different seeds or IVB with different
starting values.

The MCMC Gibbs algorithm suffers from the same problem with the dependence
between $\mathbf{W}$ and $\boldsymbol{\alpha}$, slowing down convergence.
Inspecting trace plots and the estimated inefficiency factor $IF=1+2\sum_{j=1}^{\infty}\rho_{j}$,
where $\rho_{j}$ is the autocorrelation function of the MCMC chain,
for different data sets shows that the convergence for the parameters
$\mathbf{W}$, $\mathbf{A}$ and $\boldsymbol{\lambda}$ is generally
excellent. For example, the maximum IF was less than $1.5$ across
all regression coefficients $\mathbf{W}$ for both the real data sets
when using the 3D prior, except for the head motion nuisance regressors.
With $20000$ post-burnin iterations and thinning factor $5$, this
means that we have at least $4000/1.5\approx2700$ effective samples
to base the PPMs on, and Monte Carlo standard deviations less than
$\sqrt{\frac{0.9\cdot0.1}{2700}}\approx0.0058$ for posterior probabilities
larger that $90\%$. We also looked at posterior mean maps and PPMs
for the main contrast for the different data sets and compared to
the same maps computed based only on the first $5000$, $10000$ and
$15000$ samples respectively, and saw small differences in general,
much smaller than for example when comparing to the VB maps. The spatial
hyperparameters $\boldsymbol{\alpha}$ and $\boldsymbol{\beta}$ can
however mix poorly, especially when the data are non-informative.
It would be tempting to improve the mixing using a collapsed Gibbs
sampling step for $\boldsymbol{\alpha}$ (and $\boldsymbol{\beta}$),
but this would require the computation of the precision matrix determinant
$\left|\tilde{\mathbf{B}}\right|,$ which would be too time consuming
in general. For the main parameters of interest, the $\alpha_{k}$:s
belonging to the HRF regressors, and also for the hyperparameters
belonging to the intercept and to the first AR coefficients, the convergence
rates are acceptable in general. A $1000$ iteration burnin is usually
sufficient to reach stationarity for these hyperparameters.

In the timing comparisons in the results section, all methods were
run for a long time (200 iterations for IVB, 50 iterations for SVB
and 10000 (simulated data) / 20000 (real data) iterations with thinning
factor 5 after 1000 burnin samples for MCMC (an exception was MCMC
in 3D for the face repetition data, which required 3500 burnin samples))
and we compute the time until the estimated posterior mean of $\alpha_{k}$
reaches within $1\%$ of its final values for respective algorithm.
For VB, this is the same as based on the relative error defined above,
and for MCMC this is based on the relative error of the cumulative
mean of MCMC samples. For the simulated data, this is based on all
$\alpha_{k}$, but for the real data we only consider the $\alpha_{k}$
corresponding to the intercept and HRF regressors. This is because
the convergence is sometimes extremely slow for $\alpha_{k}$:s corresponding
to the head motion and HRF temporal derivative regressors, while these
have negligible effects on the results. For the real data, we use
$\delta=10^{-8}$ and $N_{s}=100$ throughout.

The simulated data were analyzed on a computing cluster, using two
8-core (16 threads) Intel Xeon E5-2660 processors at 2.2GHz. The real
data were mainly analyzed on the same cluster, but some demanding
runs were carried out on a faster workstation with a 4-core (8 threads)
Intel Xeon E5-1620 processor at 3.5GHz. The workstation ran 46\% faster
for a large SVB estimation test and therefore the timings using this
computer were multiplied with a factor $1/0.54$ in this report, hence
the word ``approximate'' in Table \ref{tab:timingByReal}. The operating
system was Linux in both cases.

\section{Simulated and real data details\label{appendix:Data details}}

The synthetic data are simulated from the model in Section \ref{sec:Background},
with $K=5$ and $P=1$ and parameter values that are similar to those
of the pre-processed face repetition data. The design matrix is set
to have the first 4 columns equal to the standard canonical HRF regressors
from the paradigm in the face repetition data (so $T=351$) and the
fifth column corresponds to the intercept. The intercept of each voxel
is sampled i.i.d. from a $\mathcal{N}\left(900,130^{2}\right)$-distribution.
$\lambda_{n}^{-1}=100$ for each voxel and the other hyperparameters
are set as $\alpha_{1}=10^{-4},\alpha_{2}=5\cdot10^{-4},\alpha_{3}=2\cdot10^{-3},\alpha_{4}=10^{-2},\beta_{1}=10$,
which generate reasonable values of for the $\mathbf{W}$ and $\mathbf{A}$,
with varying levels of informativeness. $\mathbf{W}_{k,\cdot}$ (and
similarly $\mathbf{A}$) are sampled independently for $k\in\left\{ 1,2,3,4\right\} $
from the UGL prior using PCG with $\mathbf{B}_{data}=\mathbf{0}$
and $\mathbf{w}_{r}^{start}=\mathbf{0}$. Even though the UGL prior
is improper this works well and generates samples with mean close
to zero. Conditioned on the parameters the simulation of fMRI-data
$\mathbf{Y}$ is straightforward using the model. The data are simulated
in a big rectangular block of size $53\times63\times46$ and we follow
the SPM default to scale down the value in each voxel to get mean
100. We then use centered masks of size $10\times10\times10$, $25\times20\times20$
and $50\times50\times40$ to obtain data of size $N=10^{3}$, $10^{4}$
and $10^{5}$ that we test the methods on.

The real data from the face repetition experiment \citep{Henson2002},
previously used in \citet{pennyEtAlSpatialPrior2005} and available
at SPM's homepage \\
(\textcolor{blue}{\uline{http://www.fil.ion.ucl.ac.uk/spm/data/face\_rep/}}),
was pre-processed using the same steps as in \citet{pennyEtAlSpatialPrior2005}
using SPM12. After masking away voxels outside the brain, this results
in a data set with $N=57535$ voxels of size $3\times3\times3$ mm
and $T=351$ volumes. Using the same design matrix based on the canonical
HRF and its temporal derivative, 6 head motion parameters and an intercept,
we have $K=15$. The contrast considered is the main effect of faces
which is the mean across the four regressors corresponding to the
HRF of each condition. $P=3$ AR parameters are used for all estimations
on real data. The presented PPMs for this data set show axial slice
12, which is approximately the same region shown in \citet{pennyEtAlSpatialPrior2005}.

The real data from the visual object recognition experiment \citep{Haxby2001a,Hanson2004a,OToole2005a}
was obtained from the OpenfMRI database \\
(\textcolor{blue}{\uline{http://openfmri.org/}}). Its accession
number is ds000105 and we consider only subject001, run001. The data
was pre-processed using the same SPM pipeline as for the face repetition
data, except no slice time correction was performed (since the slice
order information is missing) neither was it normalized to a standard
brain. This data set has $N=31241$ voxels of size $3.125\times3.125\times4$
mm and $T=121$. There are 8 conditions and using the same structure
for the design matrix as for the face experiment data, we have $K=23$.
The contrast considered is the difference between seeing houses and
faces ($c_{1}=0.5$ and $c_{15}=-0.5$). The axial slice presented
in the results is number 30. 
\end{document}